\documentclass[aps,superscriptaddress,eqsecnum,nofootinbib,showpacs,preprintnumbers]{revtex4-2}
\usepackage{amsmath}
\usepackage{amssymb}
\usepackage{amsfonts,amsthm,bm}
\usepackage{color,xcolor}
\usepackage[greek,english]{babel}
\usepackage{titlesec}

\usepackage{graphicx,epsfig}
\graphicspath{{figures/}{./}}
\usepackage{float}
\usepackage{subfigure}
\usepackage{tikz}
\usepackage[pdfencoding=auto,colorlinks=true,
allcolors=blue]{hyperref}% links

\newcommand{\abs}[1]{\left| #1 \right|}
\newcommand{\be}{\begin{eqnarray}}
	\newcommand{\ee}{\end{eqnarray}}
\newcommand{\bea}{\begin{eqnarray}}
	
	\newcommand{\eea}{\end{eqnarray}}

\def\hg{\hat{g}}

\def\Ref{\ref}

\newcommand{\beq}{\begin{equation}}
\newcommand{\eeq}{\end{equation}}
\newcommand{\bseq}{\begin{subequations}}
	\newcommand{\eseq}{\end{subequations}}

\begin{document}

\title{Disformal Transition of a Black Hole to a Wormhole  in Scalar-Tensor Horndeski Theory}

\author{Nikos Chatzifotis }
\email{chatzifotisn@gmail.com} \affiliation{Physics Division,
National Technical University of Athens, 15780 Zografou Campus,
Athens, Greece.}

%\author{Kyriakos Destounis}
%\email{kyriakos.destounis@uni-tuebingen.de}
%\affiliation{Theoretical Astrophysics, IAAT, University of T$\ddot{u}$bingen, 72076 T$\ddot{u}$bingen, Germany.}

\author{Eleftherios Papantonopoulos}
\email{lpapa@central.ntua.gr} \affiliation{Physics Division,
National Technical University of Athens, 15780 Zografou Campus,
Athens, Greece.}

\author{Christoforos Vlachos}
\email{cvlach@mail.ntua.gr}
\affiliation{Physics Division,
National Technical University of Athens, 15780 Zografou Campus,
Athens, Greece.}

%\vspace{17.5cm}
%\begin{abstract}

%\end{abstract}
\vspace{6.5cm}

\begin{abstract}
We consider disformal transformations in a subclass of Horndeski theory in which a scalar field is kinetically coupled to Einstein tensor.
We apply a disformal transformation on a seed hairy black hole solution of this theory and we show that there is a transition of a black hole to a wormhole. We also show that the null energy condition is violated in the wormhole configuration and we  study the stability of the wormhole solution by calculating the time evolution of scalar perturbations in this geometry.
\end{abstract}

\maketitle

\flushbottom

\tableofcontents

\section{Introduction}

Disformal transformations were  discussed by Bekenstein \cite{Bekenstein:1992pj} in an attempt to describe gravitation with  two geometries.
The need of introducing two geometries came from the requirement that one of these describes gravitation, while the other defines the geometry in which matter describes the gravitational dynamics. This approach is very helpful if one wants to describe modified gravity theories like the scalar-tensor theories. To be consistent with various tests of general relativity (GR), to describe the dynamics of the geometry a Riemannian metric
$g_{\mu \nu}$ has to be used and to describe the gravitational dynamics one has to introduce a relation between $g_{\mu \nu}$ and the
physical geometry on which matter propagates introducing a new Riemannian metric
\begin{equation}
ds^2= \hat{g}_{\mu \nu}  dx^{\mu} dx^{\nu} \equiv \Big{(} g_{\mu \nu} A(\phi) +L^2 B(\phi) \partial_{\mu} {\phi} \partial_{\nu}{\phi}\Big{)} dx^{\mu} dx^{\nu}~,
\label{geff}
\end{equation}
where $L$ is a length scale. The physical metric $g_{\mu \nu} $ and the matter metric $\hat{g}_{\mu \nu} $ are related by a conformal and a disformal transformation.

In \cite{Bekenstein:1992pj} there was a presentation of the  physical meaning of the relation of the two metrics in \eqref{geff}. When $B=0$ a conformal transformation connects the two metrics. This transformation leaves all shapes invariant and stretches equally all spacetime directions. When $B\not= 0$ a disformal tranformation is present and its  effect is that the stretch in the direction parallel to $\partial_{\mu} \phi$ is by a different factor from that in the other spacetime directions and shapes are distorted. One has to use the metric $\hat{g}_{\mu \nu} $ from the start to see the physical context which is introduced by the disformal transformation.

Disformal transformations were used in the study of various scalar-tensor gravity theories like the Horndeski theory \cite{Horndeski:1974wa}. In \cite{Bettoni:2013diz} it was shown that disformal transformations,  for a subset of Horndeski Lagrangian, play a similar role to conformal transformations for scalar-tensor theories.
Disformal transformations were used in higher-order scalar-tensor Horndeski theories to study the stability of these theories, and the absence of ghosts \cite{Achour:2016rkg,Deffayet:2020ypa}. Cosmological perturbations were studied in these theories \cite{Tsujikawa:2014uza}, and it was shown that both scalar and tensor perturbations  on the flat isotropic cosmological background are invariant under the disformal transformation. Cosmological conformal invariance of physical observances was discussed in \cite{Domenech:2015hka}

In scalar-tensor theories the disformal coupling to matter was studied in \cite{Minamitsuji:2016hkk} and compact objects were found.
A model in a massless scalar-tensor theory was presented, and the spontaneous scalarization of slowly rotating compact objects was investigated due to the disformal coupling.  The possibility to use disformal field redefinitions to generate new hairy spherically symmetric exact solutions for quadratic
DHOST  \cite{Langlois:2017mdk,Langlois:2018dxi} theories were investigated in \cite{BenAchour:2020wiw}.  This work was further extended in \cite{Faraoni:2021gdl}  by applying a disformal transformation to  known  static or stealth scalar field seed solutions  of GR and the  disformal images
of such seeds that describe black hole horizons, wormhole throats, or horizonless geometries were discussed. In \cite{Anson:2020trg, BenAchour:2020fgy} the disformation of  the stealth Kerr black hole solution in DHOST theories was studied.

In \cite{Erices:2021uyu} the no-hair theorem was evaded and hairy black holes were found  in bi-metric  scalar-tensor theories where the two metrics were connected by conformal and disformal transformations as in  (\ref{geff}).  In this study the parameters were set as $A=L^2=1$ and $B $ was considered to be a constant independent of the scalar field $\phi$. A scalar field  coupled to the physical metric $g_{\mu\nu}$ was considered and an electromagnetic field coupled to the matter metric $\hg_{\mu\nu}$ was also introduced. Solving the field equations it was shown that the theory admits hairy black hole solutions with regular scalar field on and outside the horizon. Since the disformal factor $B$ is a coupling constant, it defines an effective cosmological constant and the spacetime can be asymptotically flat, dS or AdS. Also the thermodynamics of the hairy black hole solution was studied.

Motivated by the Bekenstein \cite{Bekenstein:1992pj} observation that the disformal transformations leads to different stretch of spacetime directions and also distorts shapes we will study the possibility of deforming a black hole to a wormhole which has a different geometry from a black hole. We will consider a hairy black hole solution of a subclass of the Horndeski theory in which the scalar hair is primary, i.e it appears in the metric function of the hairy black hole \cite{Rinaldi:2012vy}. Following \cite{Faraoni:2021gdl} we perform the disformal transformation
	\begin{equation}
		\label{3.1}
		g_{\mu\nu}\rightarrow\hat{g}_{\mu\nu}=\Omega^2(\Phi, X)g_{\mu\nu}+W(\Phi,X)\partial_\mu\Phi\partial_\nu\Phi~,
	\end{equation}
where $\Phi$ is the corresponding scalar field of the Horndeski subclass and $X=\nabla_\mu\Phi\nabla^\mu\Phi$. Then we will show that there is a critical parameter which controls the disformal transformation of the considered black hole to a wormhole. We will also study the null energy conditions and the time evolution of scalar perturbations in the shifting geometry.

The work is organized as follows. In Section \ref{sec2} the hairy black hole solution with a scalar field kinetically coupled to Einstein tensor was reviewed. In Section \ref{sec3} we discuss the generation a wormhole solution by a disformal transformation. In Section \ref{sec4} we test the null energy condition of the wormhole solution we found. In Section \ref{sec5} we study the stability of the disformed geometry by studying scalar perturbations and finally in Section \ref{sec6} we conclude.

\section{Black Hole Solution with a Scalar Field Kinetically Coupled to Einstein Tensor}
\label{sec2}

In this section we review the static hairy black hole solution found in the scalar-tensor theory in which the scalar field is kinetically coupled to Einstein tensor. This theory is a subclass of the Horndeski theory, which is a general scalar-tensor theory giving second order field equations. For the sake of completeness, we note that the full Lagrangian is given by
\begin{align}
	\label{2.1}
	\mathcal{L}&=\sum_{i=2}^{i=5}\mathcal{L}_i~,\\
	\nonumber
	\mathcal{L}_2&=K(\Phi,X)~,\\
	\nonumber
	\mathcal{L}_3&=-G_3(\Phi,X)\square\Phi~,\\
	\nonumber
	\mathcal{L}_4&=G_4(\Phi,X)R+G_{4,X}\left[(\square\Phi)^2-(\nabla_\mu\nabla_\nu\Phi)^2\right]~.\\
	\nonumber
	\mathcal{L}_5&=G_5(\Phi,X)G_{\mu\nu}\nabla^\mu\nabla^\nu\Phi-\frac{1}{6}G_{5,X}\left[(\square\Phi)^3-3\square\Phi(\nabla_\mu\nabla_\nu\Phi)^2
	+2(\nabla_\mu\nabla_\nu\Phi)^3\right]~,
\end{align}
where $X=\nabla_\mu\Phi\nabla^\mu\Phi$. The particular subset of the Horndeski theory we are considering involves non-trivial $\mathcal{L}_2=K(\Phi,X)=-\varepsilon X$ and $G_4(\Phi,X)=(8\pi)^{-1}+\frac{\eta}{2} X$ terms. Our action reads
\begin{equation}
	\label{2.2}
	S=\int d^4x \sqrt{-g}\left[\frac{R}{8\pi}-(\varepsilon g_{\mu\nu}+\eta G_{\mu\nu})\partial^{\mu}\Phi\partial^{\nu}\Phi\right],
\end{equation}
where $g_{\mu\nu}$ is a metric, $g=det(g_{\mu\nu})$, $R$ is the scalar curvature, $G_{\mu\nu}$ is the Einstein tensor, $\Phi$ is a real massless scalar field and $\eta$ is a parameter of non-minimal kinetic coupling with the dimension of length-squared. The $\varepsilon$ parameter equals $\pm 1$. In the case $\varepsilon=+1$, we have a canonical scalar field with positive kinetic term, while the case $\varepsilon=-1$ corresponds to a phantom scalar field with negative kinetic term.

Variation of the action (\Ref{2.2}) with respect to the metric $g_{\mu\nu}$ and the scalar field $\Phi$ provides the following field equations
\begin{subequations}
	\label{2.3}
	\begin{align}
		\label{2.3a}
		& G_{\mu\nu}=8\pi\big[\varepsilon T_{\mu\nu}
		+\eta \Theta_{\mu\nu}\big]~, \\
		\label{2.3b}
		&[\varepsilon g^{\mu\nu}+\eta G^{\mu\nu}]\nabla_{\mu}\nabla_\nu\Phi=0~,
	\end{align}
\end{subequations}
where
\begin{align}
	\label{2.4}
	T_{\mu\nu}&=\nabla_\mu\phi\nabla_\nu\Phi-
	\frac{1}{2}g_{\mu\nu}(\nabla\Phi)^2~, \\
	\Theta_{\mu\nu}&=-\frac{1}{2}\nabla_\mu\Phi\,\nabla_\nu\Phi\,R
	+2\nabla_\alpha\Phi\,\nabla_{(\mu}\phi R^\alpha_{\nu)}
	\nonumber\\
	&+\nabla^\alpha\Phi\,\nabla^\beta\Phi\,R_{\mu\alpha\nu\beta}
	+\nabla_\mu\nabla^\alpha\Phi\,\nabla_\nu\nabla_\alpha\Phi
	\nonumber\\
	&-\nabla_\mu\nabla_\nu\Phi\,\square\Phi-\frac{1}{2}(\nabla\Phi)^2
	G_{\mu\nu}
	\label{2.5}\\
	&+g_{\mu\nu}\big[-\frac{1}{2}\nabla^\alpha\nabla^\beta\Phi\,
	\nabla_\alpha\nabla_\beta\Phi
	+\frac{1}{2}(\square\Phi)^2 -\nabla_\alpha\Phi\,\nabla_\beta\Phi\,R^{\alpha\beta}
	\big]~. \nonumber
\end{align}
A static spherically symmetric black hole solution to the theory was found in \cite{Rinaldi:2012vy}, where it was considered that the scalar field of the theory depends only on the radial coordinate. The solution yielded the constraint that $\varepsilon\eta<0$, which led to the definition of the following parameter
\begin{equation}
	\label{2.6}
	\ell_\eta=\abs{\varepsilon\eta}^{1/2}~.
\end{equation}

In terms of the metric
\begin{equation}
	\label{2.7}
	ds^2=-g_{tt}(r)dt^2+g_{rr}(r)dr^2+g_{\theta\theta}(r)d\Omega^2~,
\end{equation}
the black hole solution corresponds to $g_{\theta\theta}(r)=r^2$ with $r\in(0,+\infty)$ and yields the following metric components
\begin{subequations}
	\label{2.8}
	\begin{align}
		\label{2.8a}
		g_{tt}(r)&=-\left[\frac{3}{4}+\frac{r^2}{12 \ell^2_\eta}-\frac{2 m}{r}+\frac{\ell_\eta}{4 	r}\arctan\left(\frac{r}{\ell_\eta}\right)\right]=-\frac{1}{4}F(r)~,\\
		\nonumber\\
		\label{2.8b}
		g_{rr}(r)&=\frac{(r^2+2\ell_\eta^2)^2}{(r^2+\ell_\eta^2)^2F(r)}~,
	\end{align}
\end{subequations}
where $F(r)=\left[3+\frac{r^2}{3 \ell^2_\eta}-\frac{8 m}{r}+\frac{\ell_\eta}{r}\arctan\left(\frac{r}{\ell_\eta}\right)\right]$,
while the scalar hair of the theory reads
\begin{equation}
	\label{2.9}
	X=-\frac{\varepsilon}{8\pi \ell_\eta^2}\frac{r^2}{r^2+\ell_\eta^2}~.
\end{equation}

An important part of the solution is that it does not fix the value of $\varepsilon$, as can be seen from (\Ref{2.9}). The only constraint of the solution is that $\varepsilon \eta<0$, which is absorbed by the kinetic coupling definition (\Ref{2.6}).
 In \cite{Rinaldi:2012vy} a canonical scalar field was considered and   the equations of motion were indeed satisfied by $\varepsilon=-1$, as long as $\varepsilon \eta<0$.
%In order to dispute this change of signs, one needs to perform the collapsing procedure of the Rinaldi black hole for $\varepsilon=-1$ and test whether %the results can be considered physical. This question is outside the scope of this paper and we leave it for future work.
However, since $\Phi$ is solely dependent on the radial coordinate, then the vector $\partial^\mu\Phi$ is spacelike, which yields the result that $\varepsilon=-1$, i.e. the considered black hole is indeed generated by a phantom scalar field. For the sake of a general analysis on this, we chose not to fix the value of $\varepsilon$.
The non-minimal coupling constant sources an asymptotic AdS spacetime therefore, this solution  reproduces the Schwarzschild black hole in the limit of $l_\eta\rightarrow+\infty$, therefore the geometry can be understood as a hairy black hole generalization of the Schwarzschild spacetime with effective AdS-asymptotics, when the spin-0 degree of freedom also acquires dynamics from the kinetic mixing with the graviton, i.e. the $G^{\mu\nu}\partial_\mu\Phi\partial_\nu\Phi$ term.

At the linearized
level, it was shown in \cite{Vlachos:2021weq} that because of the  asymptotic AdS-like boundary, which  serves as a perfect reflector for
incident scalar waves generated by  scalar perturbations of a test scalar field, an effective potential is generated
outside the horizon of the black hole trapping the incident scalar waves. Then it was found that the ringdown signal of
the black exhibit successively damped echoes, indicating the stability of the black hole. Then calculating gravitational perturbations it was shown \cite{Chatzifotis:2021pak} that this hairy compact object is stable.

\section{Generating a wormhole solution by a disformal transformation}
\label{sec3}

Motivated by the work in \cite{Faraoni:2021gdl}, we  propose a way to create a geometry that can interpolate between black hole, regular black hole, one-way wormhole,  i.e. a wormhole geometry with the throat radius being equal to an event horizon radius,  and two-way wormhole solution. Performing the disformal transformation (\ref{3.1}) to the metric (\ref{2.7}) we get
\begin{equation}
	\label{3}
	d\hat{s}^2=-\Omega^2(\Phi,X)\frac{1}{4}F(r)dt^2+\frac{\Omega^2(\Phi,X)+W(\Phi,X)X}{\frac{(r^2+l_\eta^2)^2}{(r^2+2l_\eta^2)^2}F(r)}dr^2
+\Omega^2(\Phi,X)r^2d\Omega^2~.
\end{equation}
We note that the conditions for the disformal transformation to be an invertible map reads \cite{Faraoni:2021gdl}
\begin{align}
	\label{3.2}
	&\Omega\neq0~,\\
	\label{3.3}
	&\Omega^2-X(\Omega^2)_{X}-X^2W_X\neq0~.
\end{align}

The presence or absence of black hole horizons can be assessed by studying the norm of the Kodama vector, $\nabla_\mu R\nabla^\mu R$, where $R$ is the $g_{\theta\theta}$ component of the metric. In \cite{Faraoni:2021gdl} it was argued that a black hole horizon corresponds to single positive root, while a double root corresponds to a wormhole throat. The norm of the Kodama vector associated with the metric (\Ref{3}) reads
\begin{equation}
	\label{3.4}
	\nabla_\mu R\nabla^\mu R=4\frac{\frac{(r^2+l_\eta^2)^2}{(r^2+2l_\eta^2)^2}F(r)}{\Omega^2(\Phi,X)+W(\Phi,X)X}\left(\Omega_\Phi(\Phi)' r^2+\Omega_X(X)' r^2+\Omega^2r\right)^2~,
\end{equation}
where prime denotes differentiation with respect to the radial coordinate $r$. If any root to the second term of the right hand side of the above equation exists, then this root corresponds to a wormhole throat. However, since we wish our resulting geometry to be able to interpolate between different compact objects, this method will not do. Instead, we would like to focus on the denominator of (\Ref{3.4}) and choose an appropriate $W$ function. Also, in order for the black hole geometry to be able to be recovered, it will prove useful to set the conformal factor $\Omega=1$. Naturally, since this choice does not cancel out the null hypersurface of the event horizon, the existence of the event horizon needs to become dependent on a new parameter, which we will introduce. We recall that, for the (\ref{2.8a}), (\ref{2.8b}) solution, $X$ reads
\begin{equation}
	\label{3.5}
	X=\frac{-\varepsilon}{8\pi l^2_\eta}\frac{r^2}{r^2+l^2_\eta}~,
\end{equation}
which straightforwardly leads to
\begin{equation}
	\label{3.6}
	r^2=\frac{-8\pi l_\eta^4 X}{8\pi l_\eta^2 X+\varepsilon}~.
\end{equation}
Keeping the shift symmetry of the theory, we can set the dependence of $W$ to be $W=W(X)$. Our goal is to modify the $g_{rr}$ component of the metric to contain the term $\frac{r^2}{r^2-a^2}$, where $a$ is a new parameter with dimensions of length squared. According to (\Ref{3.6}), the appropriate $W$ function can easily be found to be
\begin{equation}
	\label{3.7}
	W=-\frac{1}{X}+\frac{8\pi l_\eta^4 X}{8\pi l_\eta^4 X+a^2(8\pi l_\eta^2 X+\varepsilon)}~,
\end{equation}
which yields the following metric
\begin{equation}
	\label{3.8}
	d\tilde{s}^2=-\frac{1}{4}F(r)dt^2+\frac{dr^2}{\frac{(r^2-a^2)}{r^2}\frac{(r^2+l_\eta^2)^2}{(r^2+2l_\eta^2)^2}F(r)}+r^2d\Omega^2~.
\end{equation}
This metric describes a wormhole as long as the new parameter $a$ is larger than the event horizon of the underlying black hole solution. In order to see that more concretely, let us perform a coordinate transformation to cover the region $r>a$ two times. This is easily done by the choice $x^2=r^2-a^2,\,\,\, x\in\mathbb{R}$, which yields $\displaystyle\left(\frac{dr}{dx}\right)^2=\frac{r^2-a^2}{r^2}$ and our final metric reads
\begin{equation}
	\label{3.9}
	d\tilde{s}^2=-\frac{1}{4}F(\sqrt{x^2+a^2})dt^2+\frac{dx^2}{\frac{(x^2+a^2+l_\eta^2)^2}{(x^2+a^2+2l_\eta^2)^2}F(\sqrt{x^2+a^2})}+(x^2+a^2)d\Omega^2~.
\end{equation}
This metric indeed interpolates between a black hole, regular black hole, one-way wormhole and two-way wormhole solution as $a$ grows from $0$ to values larger than $r_h$, where $r_h$ is the event horizon of the underlying black hole solution. We note that this is possible because the $g_{tt}$ component, $F(r)$ is a monotonically increasing function and we can use the disformal transformation to enforce that the metric components are positive definite. In the case of $a>r_h$, the $x=0$ hypersurface describes the wormhole throat.  On more mathematical grounds this result can be understood as follows. Following \cite{Hochberg:1997wp}, a traversable wormhole throat, $\Sigma$, is a two-dimensional hypersurface of minimal area taken in one of the constant-time spatial slices.  In order to compute the area, $A_\Sigma$ we make use of the equation
 \begin{equation}
 	\label{3.10}
 	A_{\Sigma}=\int \sqrt{g^{(2)}}d^2x~,
 \end{equation}
under normal Gaussian coordinates $g_{ij}dx^idx^j=dn^2+g_{ab}dx^adx^b$.
Then, in order for $\Sigma$ to be a traversable wormhole throat, the minimal area condition needs to be satisfied, i.e.
\begin{align}
	\label{3.11}
	\delta A_{\Sigma}=\int \partial_n \sqrt{g^{(2)}}d^2x=-\int\sqrt{g^{(2)}}Tr(K)\delta n d^2x=0~,\\
	\label{3.12}
	\delta^2A_{\Sigma}=-\int\sqrt{g^{(2)}}(\partial_nTr(K)-Tr(K)^2)\delta n\delta n d^2x\geq0~,
\end{align}	
where $K_{ab}=-\frac{1}{2}\partial_n g_{ab}$ denotes the extrinsic curvature of $\Sigma$.	
Indeed, using the disformed line element, (\Ref{3.9}), we find after some  algebra that both the extremality condition,
	\begin{equation}
		\label{3.13}
		Tr(K)=-\frac{1}{2}g^{ab}\partial_n g_{ab}=-\frac{1(1+\sin^2\theta)}{2(x^2+a^2)}\partial_x (x^2+a^2)\frac{1}{\sqrt{g_{xx}}}\overset{x=0}{=}0~,
	\end{equation}
and the minimal area condition,
	\begin{equation}
		\label{3.14}
		-\partial_n Tr(K)=-\partial_x Tr(x)\frac{1}{\sqrt{g_{xx}}}\overset{x=0}{=}\frac{4(a^2+\ell_\eta^2)^2(1+\sin^2\theta)}{(a^3+2a\ell_\eta^2)^2}g_{tt}(0)>0~,
	\end{equation}
are satisfied as long as the $a$ parameter is greater then the horizon radius.

One final note is that, similarly to the (\ref{2.8a}), (\ref{2.8b})  black hole, which yields the Schwarzschild geometry in the limit where the kinetic coupling becomes infinite, this solution yields the bouncing solution of \cite{Simpson:2018tsi}
\begin{equation}
	\label{3.15}
	ds^2=-\left(1-\frac{2m}{\sqrt{x^2+a^2}}\right)dt^2+\frac{dr^2}{\left(1-\frac{2m}{\sqrt{x^2+a^2}}\right)}+(x^2+a^2)d\Omega^2~,
\end{equation}
 in the limit of $l_\eta\rightarrow\infty$, which has recently been the subject of intensive study \cite{Stuchlik:2021tcn,Chakrabarti:2021gqa,Domenech:2019syf,Franzin:2021vnj,Islam:2021ful,Tsukamoto:2020bjm,Lobo:2020ffi,Churilova:2019cyt}.

 Having found this solution  an interesting question is to find the action which can potentially source these type of solutions. In this action, the seed metric tensor, $g_{\mu\nu}$, should be expressed  in terms of the disformed $\hat{g}_{\mu\nu}$. However, the metric $\hat{g}_{\mu\nu}$ contains terms that are contracted with the original seed metric $g_{\mu\nu}$, namely the term (\Ref{3.7}), and the resulting action would contain coupled terms of the two metrics. A way out of this is to analytically express the original seed metric $g_{\mu\nu}$ in terms of the disformed one, i.e. to find the inverse transformation which is a difficult task.

%A second problem that occurs is that, under the proposed disformal transformation, one needs to keep the value of the scalar field that sources the %bouncing solution fixed. In the action level, this could easily be done by using an appropriate Lagrangian multiplier. One would argue that this procedure %would be on the borderline of fine-tuning, but this is a wrong assumption.} The bouncing solution we are proposing is derived from the original subclass %of the Horndeski geometry, which is being sourced by the Galileon field. As such, we indeed need to keep the solution of the Galileon field fixed, in %order to get a bouncing solution.
Let us now discuss the general setup of our approach. The bouncing solution should be a solution to a general action in Einstein frame of the form
\begin{equation}
	\label{3.16}
	S=\int d^4x \sqrt{-\hat{g}}\frac{\hat{R}}{8\pi}+\mathit{\hat{L}}_m~,
\end{equation}
where $\mathit{\hat{L}}_m$ denotes an arbitrary matter Lagrangian. It is clear that the equations of motion that (\Ref{3.16}) yields are of the known form
\begin{equation}
	\label{3.17}
	\hat{G}_{\mu\nu}=8\pi \hat{T}_{\mu\nu}~.
\end{equation}
Then, we can readily express the left hand side of (\Ref{3.17}) in terms of the original seed metric and find the form of the stress energy tensor $\hat{T}_{\mu\nu}$, which sources the bouncing solution. In particular, we find as we explain in detail in the Appendix, that the connection receives a term attributed to the disformal transformation,
\begin{equation}
	\label{3.18}
	D^{\mu}_{\,\,\,\sigma\nu}=\frac{1}{2(1+WX)}\left[\nabla^{\mu}\Phi(\nabla_\nu W \nabla_\sigma \Phi+ \nabla_\sigma W \nabla_\nu \Phi + 2 W \nabla_\nu\nabla_\sigma\Phi)+ \nabla_\sigma\Phi\nabla_\nu\Phi(W \nabla^\mu\Phi \nabla^\beta \Phi \nabla_\beta W -\nabla^\mu W(1+W X))\right]~,
\end{equation}
while the left hand side of (\Ref{3.17}) reads
\begin{equation}
	\label{3.19}
	\hat{G}_{\mu\nu}(\hat{g})=G_{\mu\nu}(g)+S_{\mu\nu}(g,\partial_\mu\Phi)~.
\end{equation}
We  stress  that the right hand side of (\Ref{3.19}) contains tensors derived in terms of the original seed metric and the sourcing Galileon field. As such, if we had an analytical expression of the inverse transformation, one could readily find the new stress energy tensor $\hat{T}_{\mu\nu}$
\begin{equation}
	\label{3.20}
	\hat{T}_{\mu\nu}(g(\hat{g}),\partial_\mu\Phi)=\frac{G_{\mu\nu}(g(\hat{g}))}{8\pi}+\frac{S_{\mu\nu}(g(\hat{g}),\partial_\mu\Phi)}{8\pi}\overset{(\Ref{2.3a})}{=}\left[\varepsilon T_{\mu\nu}(g(\hat{g}),\partial_\mu\Phi)+\eta \Theta_{\mu\nu}(g(\hat{g}),\partial_\mu\Phi)\right]+\frac{S_{\mu\nu}(g(\hat{g}),\partial_\mu\Phi)}{8\pi}~.
\end{equation}
The above expression  shows us that it is possible to derive the bouncing solution from the original black hole solution by upgrading the preexisting stress energy tensor with the addition of a new tensor term $S_{\mu\nu}$.

 From the above relation of the energy-momentum tensor, (\Ref{3.20}), one can deduce that the wormhole solution we find using the disformal transformation is indeed just a deformation of the original black hole geometry we considered. In accordance with our discussion on Bekenstein's work \cite{Bekenstein:1992pj} in the Introduction, the gravitation is described by the disformed metric $\hat{g}_{\mu \nu} $, while the stress-energy tensor is described by the original physical seed metric $g_{\mu \nu} $ coupled to the matter field of the theory. As such, we can express the corresponding action in a qualitative manner
\begin{equation}
	\label{3.21}
		S=\int d^4x \sqrt{-\hat{g}}\left[\frac{\hat{R}(\hat{g})}{8\pi}+\frac{\sqrt{-g}}{\sqrt{-\hat{g}}}\left(\mathit{L}_{NMDC}(g,\partial_\mu\Phi)
+\mathit{L}_{D}(g,\partial_\mu\Phi)+\mathit{L}_C(g,\partial_\mu\Phi)\right)~\right]~,
\end{equation}
where $\mathit{L}_{NMDC}$ is the matter Lagrangian of the original theory, $\mathit{L}_D$ is a matter Lagrangian, whose variation yields the last term of (\Ref{3.20}) and $\mathit{L}_C$ yields the constraint of the matter field to the original solution via an appropriate Lagrangian multiplier. Using the associated relation between the two metrics, we can express the disformed volume form
\begin{equation}
	\label{3.22}
	\sqrt{-\hat{g}}d^4x=\Omega^3(\Omega^2+W X)^{1/2}\sqrt{-g}d^4x~,
\end{equation}
in terms of the seed geometry via Sylvester's determinant theorem.

\section{Testing of the energy conditions}
\label{sec4}

Under the simple expression of (\Ref{3.17}), it is possible to test the possible violations of the null energy conditions as the $a$ parameter is increasing and the solutions shifts from the original black hole to the two-way wormhole. In order to simplify our calculations, we consider that the event horizon of the original black hole is the null hypersurface $r=1$. Therefore, we can express the mass parameter of the black hole $m$ in terms of the kinetic coupling $l_\eta$ as follows
\begin{equation}
	\label{4.1}
	m=\frac{1}{24}\left(9+\frac{1}{l_\eta^2}+3l_\eta\arctan\left[\frac{1}{l_\eta}\right]\right)~,
\end{equation}
and consequently shrink the parameter space of our geometry. As a test, we note that, as expected, the mass parameter limits to $\displaystyle\frac{1}{2}$ as $l_\eta\rightarrow\infty$, which is the Schwarzschild black hole limit. In the figures below, we present the null energy condition violations for different values of the throat parameter $a$.
\begin{figure}[h!]
	\centering
	\includegraphics[scale=0.5]{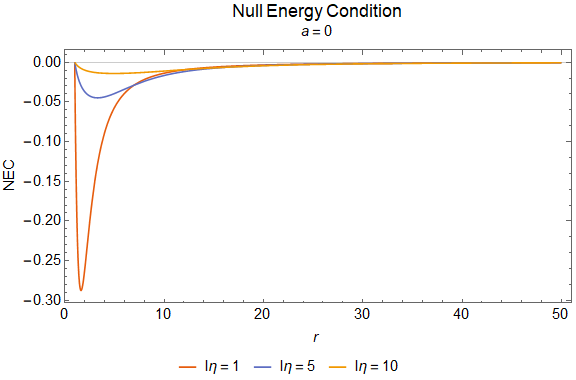}
	\caption{The violation of the null energy conditions in the black hole case for different values of the kinetic coupling.} \label{fig:BH}
\end{figure}
\begin{figure}[h!]
	\centering
	\includegraphics[scale=0.4]{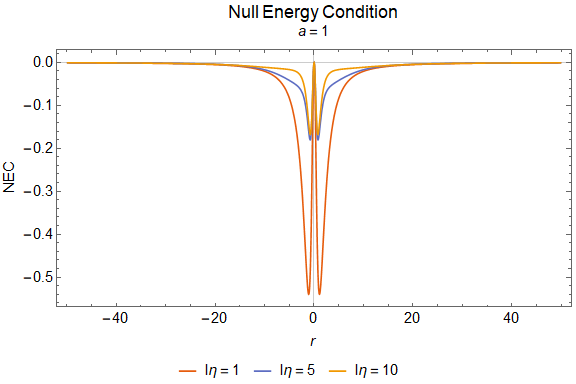}\hspace{0.2 cm}
	\includegraphics[scale=0.4]{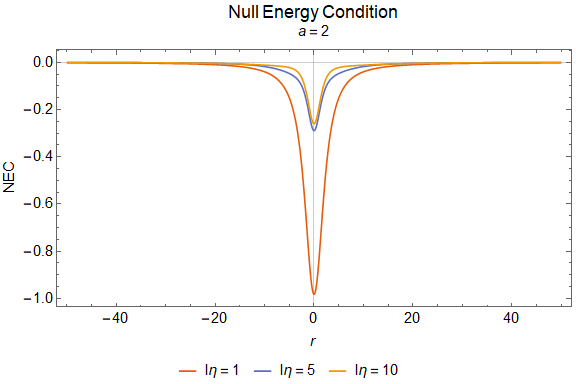}
	\caption{Left: The violation of the null energy conditions in the one-way wormhole case for different values of the kinetic coupling.
	Right: The violation of the null energy conditions in the two-way wormhole case for different values of the kinetic coupling.} \label{fig:wh}
\end{figure}

 We first discuss the  null energy condition (NEC) in the case of the black hole when $a$ is zero. As can be seen in Fig.~\Ref{fig:BH}, near the horizon the NEC is violated and the violation is stronger for small kinetic coupling $l_\eta$, while away from the horizon of the black hole the NEC is restored. This beheviour of the NEC can be understood because at small kinetic coupling $l_\eta$ the gravitational attraction is stronger and the kinetic energy of the scalar field should balance this attraction force for a hairy black hole to be formed. Away from the black hole horizon the hairy black hole with a primary scalar charge is formed, the violation of the NEC is restored and as it was shown in \cite{Chatzifotis:2021pak} the calculation of axial tensor perturbations shows that the hairy black hole is stable. We also note that during this process, the physical requirement that the local energy density as measured by a static observer is a positive definite quantity is satisfied.

In the case that $a$ is non-zero, we can see from Fig.~\Ref{fig:wh} that as the  throat radius grows the violation of the NEC is stronger, as it is expected. This means that the $S_{\mu\nu}$ tensor (\Ref{A.12}) we derived in the Appendix,  strongly violates the null energy conditions, but the exotic matter is contained in the wormhole vicinity. In the next Section we  test the perturbations of the two-way wormhole solution under scalar perturbation and compare the results to those of the original black hole case. We  expect the wormhole solution for low values of $\ell_\eta$ to be unstable, since the sourcing matter becomes highly exotic in this case.

\section{Evolution of scalar perturbations on  the wormhole geometry}
\label{sec5}

In this Section we study the time evolution of scalar perturbations in our shifting geometry. This entails the study of the propagation of a test scalar field in the vicinity of the compact object under question. In particular, our goal is to extract the Regge-Wheeler potential of the scalar perturbations for different values of the throat radius $a$ and use the finite difference method, first proposed in \cite{Gundlach:1993tn}, to numerically solve the wave equation. In order to keep the analysis as general as possible, we shall keep the metric components in the form of
		\begin{equation}
			\label{5.1}
			ds^2=-f(x)dt^2+g(x)dx^2+\rho^2(x)d\Omega^2~,
		\end{equation}	
	where in the black hole case, the radial coordinate $x=r$ and is positive definite, while the $g_{\theta\theta}$ component is equal to $r^2$, as is the usual ansatz for spherically symmetric spacetimes. In the wormhole case, $x$ takes values in the range $(-\infty,+\infty)$ and the $g_{\theta\theta}$ component is equal to $x^2+a^2$.
The Klein-Gordon equation of motion for a test massless scalar field $\Psi$ in a spherically symmetric curved background reads
\begin{equation}
	\label{5.2}
	\frac{1}{\sqrt{-g}}\partial_\mu\left[
	\sqrt{-g}g^{\mu\nu}\partial_\nu\Psi\right]=0~.
\end{equation}
We choose the ansatz of $\Psi(t,x,\theta,\phi)=R(x)
Y^l_m(\theta,\phi)e^{-iwt}$ to disentangle the radial, angular and temporal parts of the field. Under this ansatz the Klein-Gordon equation of motion is reduced to the
\begin{equation}
	\label{5.3}
	h(x)\partial_{x}\left[\rho^2(x)h(x)\partial_{x}R(x)\right]+[w^2-l(l+1)f(x)]R(x)=0~,
\end{equation}
where we set $h(r)=\frac{\sqrt{f(x)}}{\sqrt{g(x)}}$ and the $l$ parameter denotes the orbital state number of the test scalar field.
Performing the transformation to the tortoise coordinate $x^*$, $\displaystyle dx^*=\frac{dx}{h(x)}$,
we simplify the above equation to
\begin{equation}
	\label{5.4}
	\partial_{x^*}\left[\rho^2(x)\partial_{x^*}R(x)\right]+[w^2-l(l+1)f(x)]R(x)=0~.
\end{equation}
Finally, performing the substitution $R(x)=\frac{\psi(x)}{\rho(x)}$, the radial equation takes the following form
\begin{equation}
	\label{5.5}
	\frac{\partial^2 \psi(x)}{\partial x^{*2}}+\left[w^2-V_{RW}(x)\right]\psi(x)=0~.
\end{equation}
As such, the explicit form of the Regge-Wheeler potential reads
\begin{equation}
	\label{5.6}
	V_{RW}=l(l+1)\frac{f(x)}{\rho^2(x)}+\frac{2f(x)g(x)\frac{\partial^2\rho(x)}{\partial x^2}+g(x)\frac{\partial f(x)}{\partial x}\frac{\partial \rho(x)}{\partial x}-f(x)\frac{\partial g(x)}{\partial x}\frac{\partial \rho(x)}{\partial x}}{2g(x)\rho^2(x)}~.
\end{equation}
Having the Regge-Wheeler potential we can use the time-domain integration method \cite{Gundlach:1993tn} to calculate the temporal response of linear massless scalar field perturbations on the wormhole geometry. We will give a brief outline of the method. By defining $\psi(x_\ast,t)=\psi(i\Delta x_\ast,j\Delta t)=\psi_{i,j}$, $V(x(x_*))=V(x_\ast,t)=V(i\Delta x_\ast,j\Delta t)=V_{i,j}$, equation (\ref{5.5})  takes the form
\begin{align}
    \frac{\psi_{i+1,j} - 2\psi_{i,j} + \psi_{i-1,j} }{\Delta x^2_\ast} - \frac{ \psi_{i,j+1} - 2\psi_{i,j} + \psi_{i,j-1} }{\Delta t^2} - V_i \psi_{i,j} = 0\,.
\end{align}
Then, by using as initial condition a Gaussian wave-packet of the form $\psi(x_\ast,t) = \exp\left[ -\frac{(x_\ast-c)^2}{2\sigma^2} \right]$ and $\psi(x_\ast,t<0) = 0$, where $c$ and $\sigma$ correspond to the median and width of the wave-packet, we can derive the time evolution of the scalar field $\psi$ by
\begin{align}
    \psi_{i,j+1} = -\psi_{i,j-1} + \left(\frac{\Delta t}{\Delta x_\ast}\right)^2\left( \psi_{i+1,j} + \psi_{i-1,j} \right) + \left( 2 - 2\left(\frac{\Delta t}{\Delta x_\ast}\right)^2 - V_i \Delta t^2 \right) \psi_{i,j}
    \label{psi-evolution}\,,
\end{align}
where the Von Neumann stability condition requires that $\frac{\Delta t}{\Delta x_\ast} < 1$.
The effective potential is positive and vanishes at the null hypersurface of the one-way wormhole (but not at the throat of the two way wormhole), however, it diverges as one approaches the asymptotic spatial infinity for both compact objects. This requires that $\psi$ should vanish at infinity, which corresponds to reflective boundary conditions.  To calculate the precise values of the potential $V_i$, we integrate numerically the equation for the tortoise coordinate and then solve with respect to the corresponding radial coordinate. Various convergence tests were performed throughout our numerical evolution, with different integration steps and precision, to reassure the validity of our ringdown profiles.

By applying the numerical procedure outlined above, we calculate the temporal response of linear massless scalar field perturbations on the discussed  wormhole solutions. In both cases, the perturbation response is obtained at a position arbitrarily close to the throat.
%------------------------------
\subsection{Two-Way Wormhole}
\begin{figure}[]
	\centering
	\includegraphics[scale=0.485]{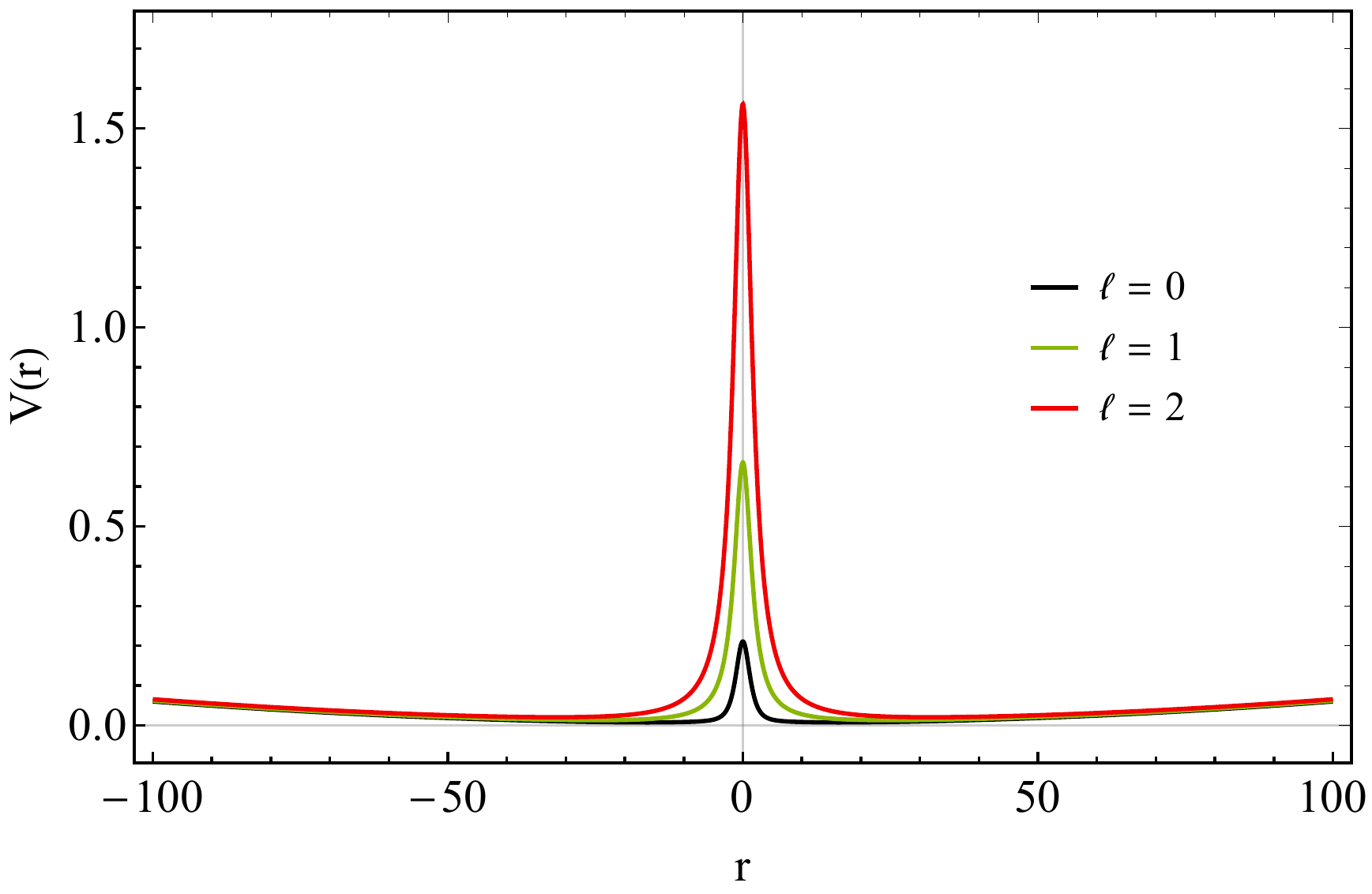}\hspace{0.2 cm}
	\includegraphics[scale=0.5]{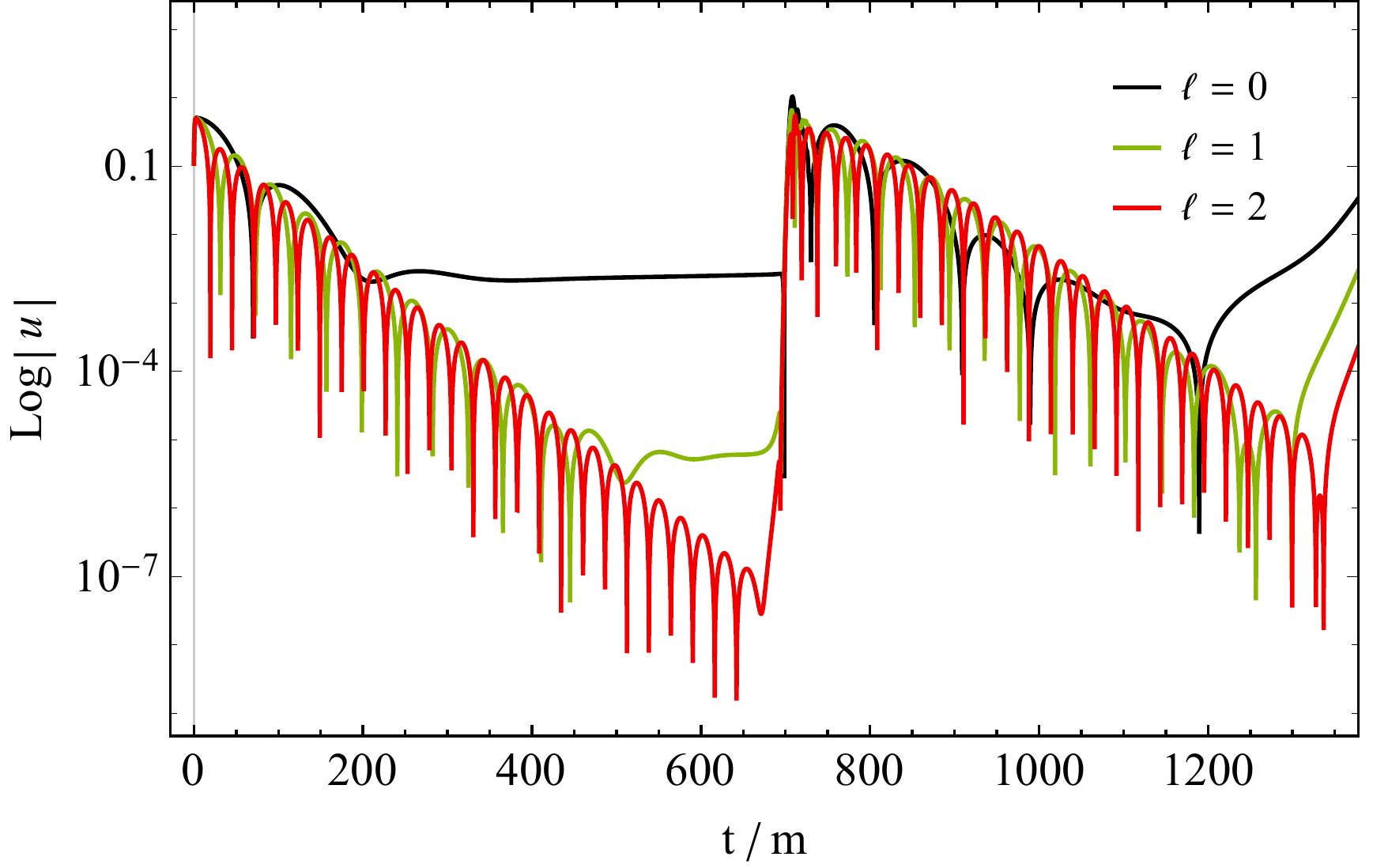}
	\caption{Effective potential (left) and time evolution (right) of scalar perturbations, with varying $\ell$, of the two-way wormhole with  $\ell_\eta = 10$, $\alpha = 2$ and $m = 0.1$. }
%Higher angular momenta correspond to larger potential barriers at the throat of the wormhole. Thus when the compact object is perturbed, more energy is %carried away from the photon sphere (PS) leading to more oscillatory signals.}
\label{fig:1}
\end{figure}

%------------------------------
In Fig.~\ref{fig:1} we demonstrate the behavior of a test scalar field as it propagates in the two-way wormhole background \eqref{3.9}. The parameters are tuned in such a way that $\alpha>r_h$. The most obvious effect is the emergence of echoes following the initial quasinormal ringdown. This pattern becomes more evident for any increment of the angular momentum $\ell$, due to the fact that more energy is carried away from the photon sphere (PS) when perturbed, resulting in a more oscillatory signal.

\begin{figure}[]
	\centering
	\includegraphics[scale=0.485]{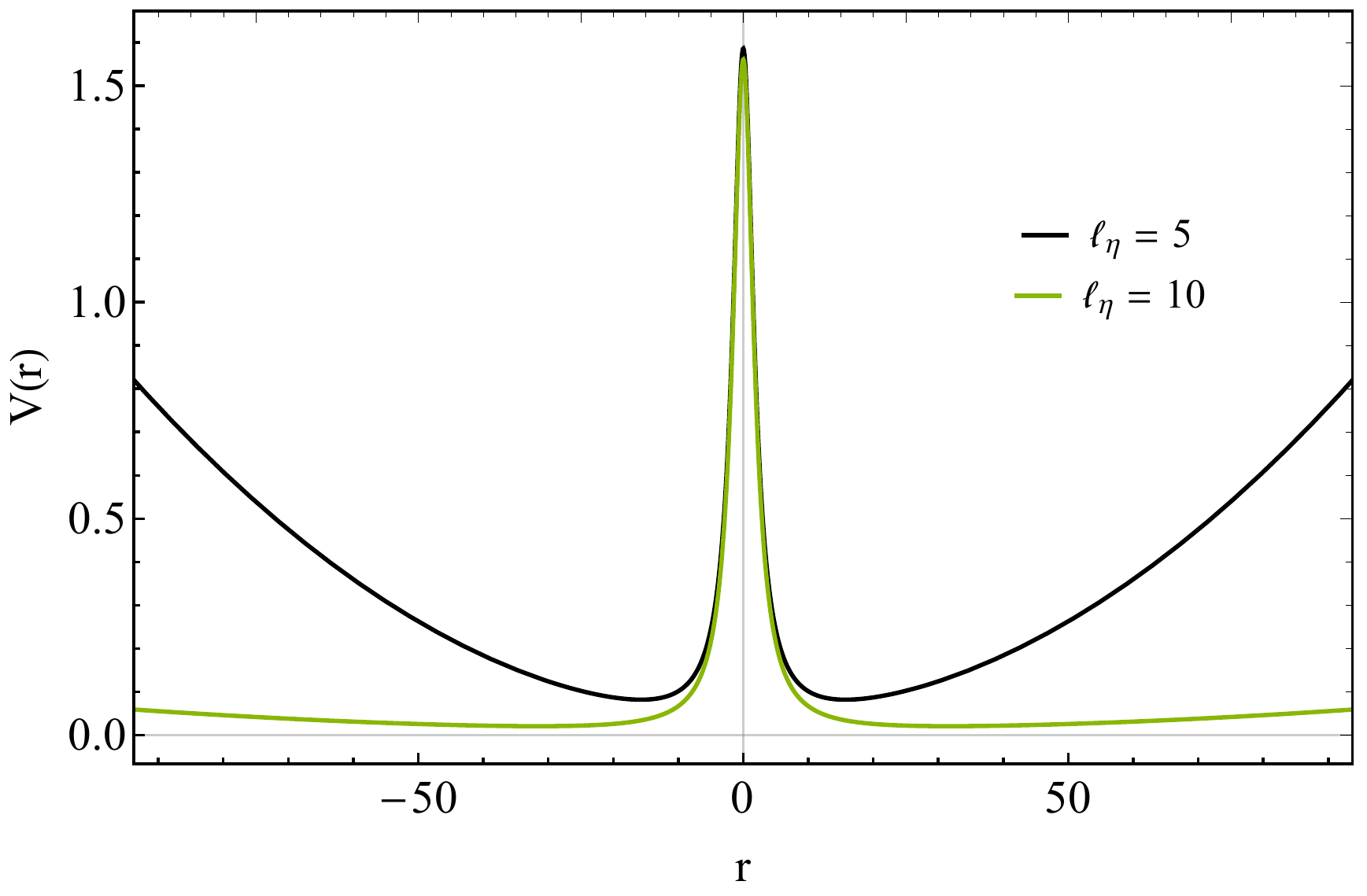}\hspace{0.2 cm}
	\includegraphics[scale=0.5]{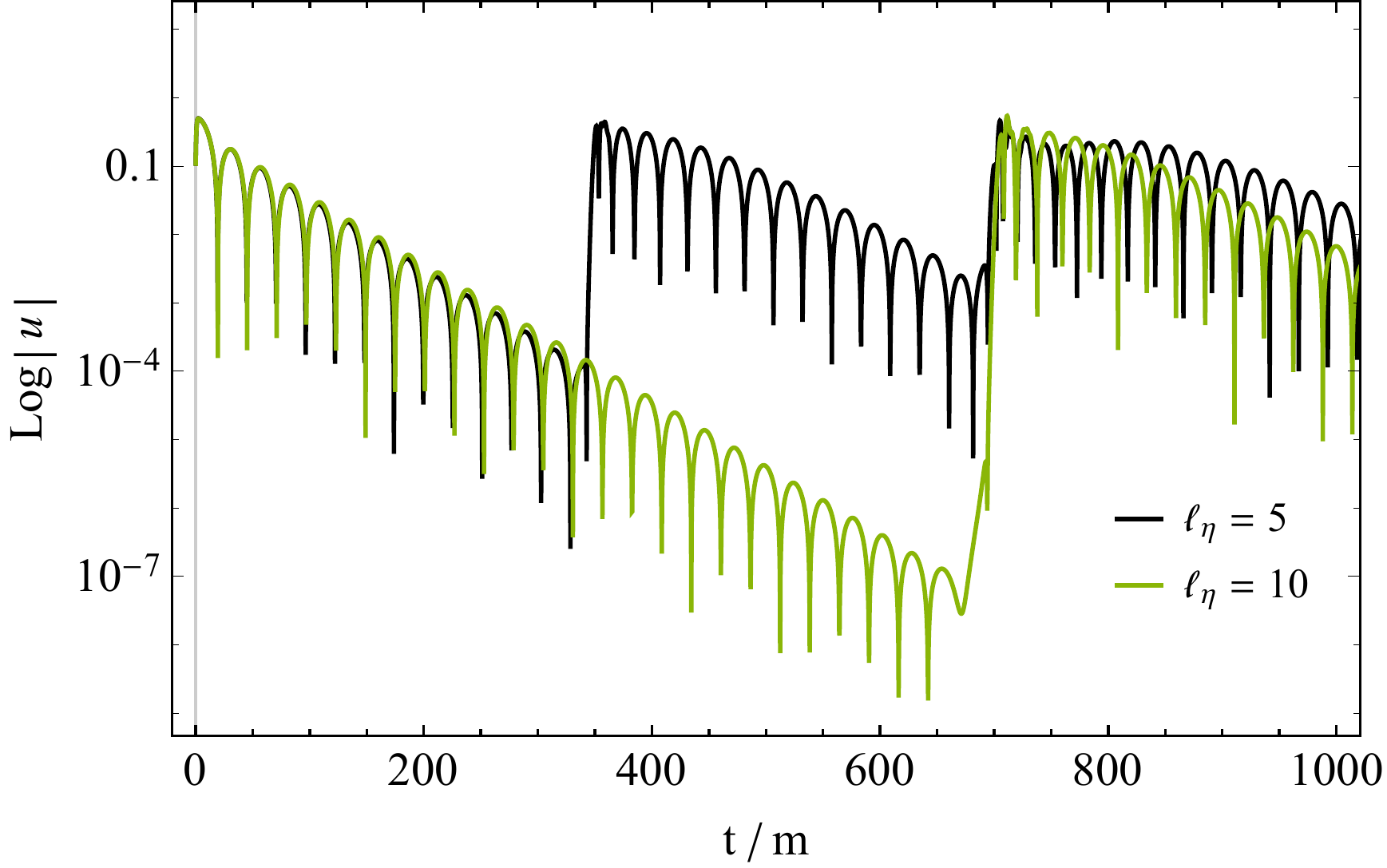}
	\caption{Effective potential (left) and time evolution (right) of scalar perturbations, with varying $\ell_\eta$, of the two-way wormhole with  $\ell = 2$, $\alpha = 2$ and $m = 0.1$. }
%When $\ell_\eta$ increases the effective AdS boundary moves further away from the throat. As a result the perturbations reflected off the PS have to %travel a greater distance before they reach the AdS boundary and return to re-perturb the PS.}
\label{fig:2}
\end{figure}

In Fig.~\ref{fig:2} we fixed the angular momentum and we vary the kinetic coupling $\ell_\eta$. The increase of $\ell_\eta$, which acts as an effective cosmological constant giving an AdS spacetime, moves the effective AdS boundary further away from the throat. As a result the perturbations reflected off the PS have to travel a greater distance before they reach the reflective AdS boundary and return to re-perturb the PS. Hence, any increment of $\ell_\eta$ results in a delay of the echoes.

\begin{figure}[]
	\centering
	\includegraphics[scale=0.485]{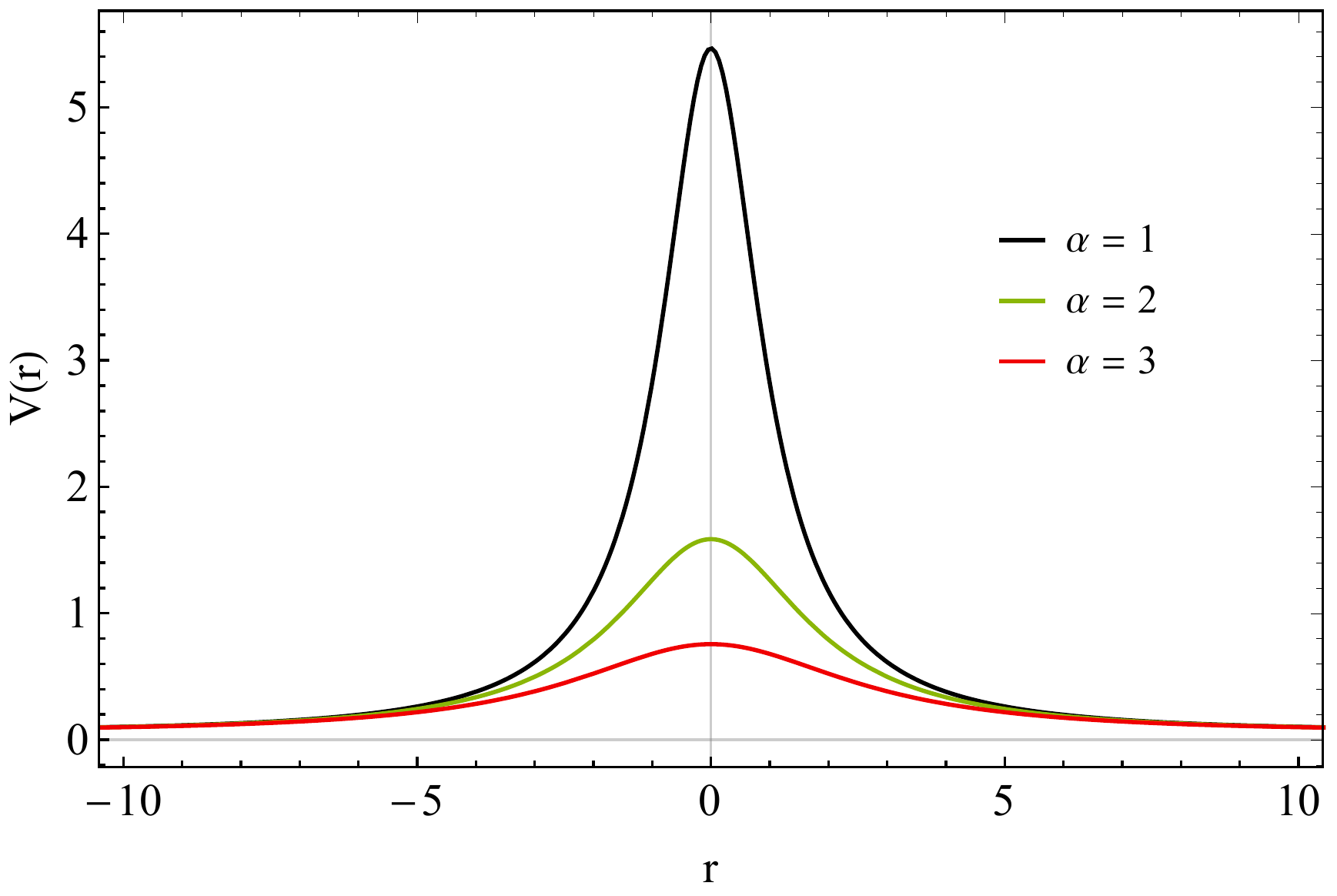}\hspace{0.2 cm}
	\includegraphics[scale=0.5]{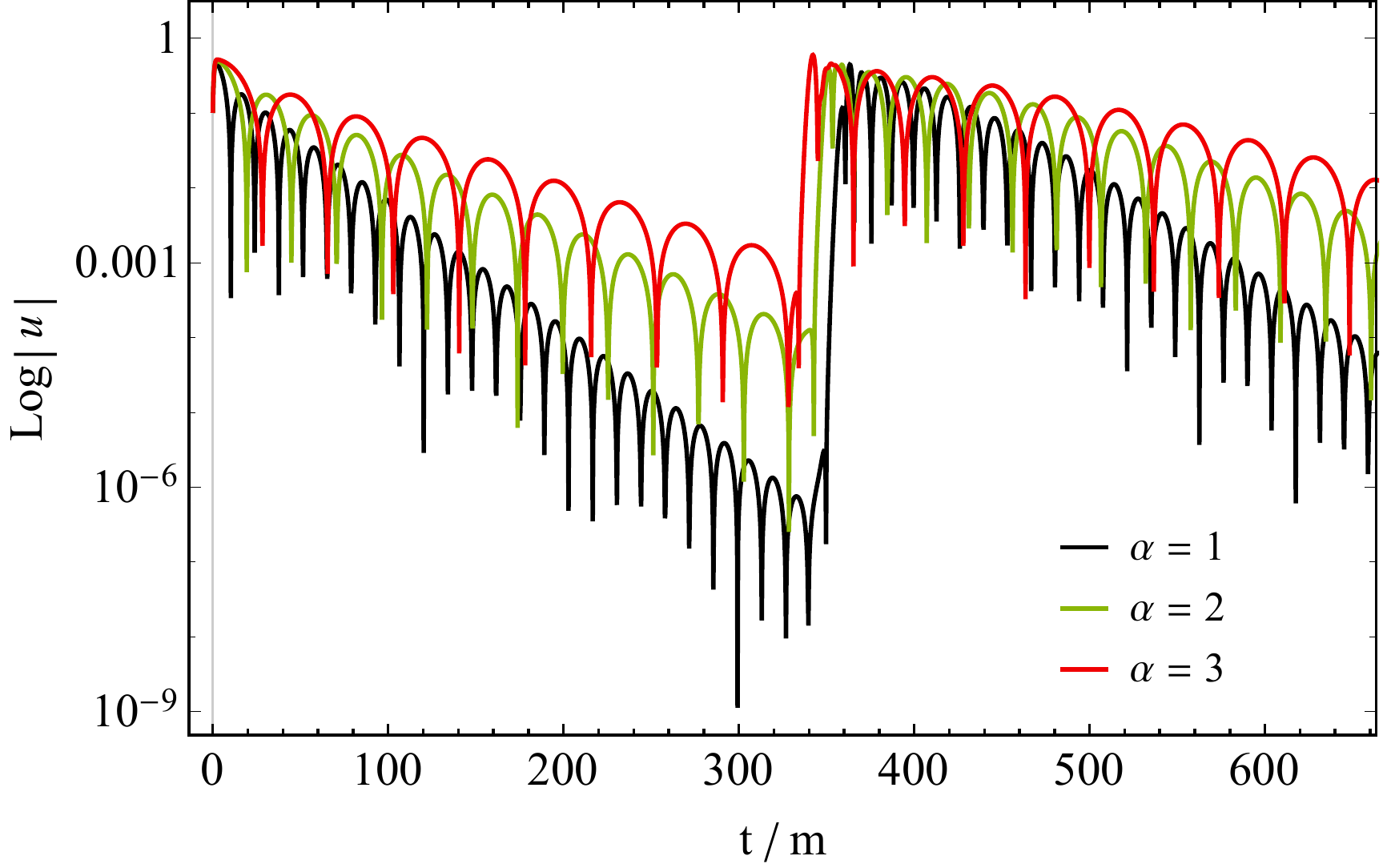}
	\caption{Effective potential (left) and time evolution (right) of scalar perturbations, with varying $\alpha$, of the two-way wormhole with  $\ell = 2$, $\ell_\eta = 5$ and $m = 0.1$.%Notice that the throat size $\alpha$ directly affects both the oscillation ($\tau=1/\omega_r$) and the damping time scale ($\tau=1/\omega_i$) of our
 % signal.
  } \label{fig:3}
\end{figure}

In Fig.~\ref{fig:3} the effect of the throat size $\alpha$ is illustrated. We can see that both the oscillation ($\tau_r=1/\omega_r$) and the damping time ($\tau_i=1/\omega_i$) of our  signal are effected. Evidently, this behavior stems from the shape of the effective potential. As $\alpha$ increases less energy is carried away from the PS leading to an increase in the oscillation time. Moreover, the slope of the potential lessens, which causes the increase in the damping time of the signal.

Finally, it is important to note that the amplitude of the echoes does not decrease with time, a behavior which indicates the absence of energy dissipation. Our test field travels through the throat and into the second Universe only to be reflected back from the second AdS boundary. This absence of dissipation is an indication that our compact object may possess normal modes of oscillations similar to the ones found in
\cite{Gundlach:1993tp, Leaver:1986gd, Correa:2008nq, Evnin:2015gma, Evnin:2017vpc, Fierro:2018rna, Anabalon:2019lzc, Shaikh:2021yux}. However, a mode decomposition of the probe field to calculate these modes is a rather challenging task due to the complicated form of the metric components and is out of the scope of the present study.
%----------------------------------------------
%-------------------------------------------------
%------------------------------------------------------
%------------------------------------------------------
%\begin{figure}[H]\centering\includegraphics[scale=0.479]{Two_way_plots/Twoway_a1leta5l2_m020304_Veff.pdf}\hspace{0.2 cm}\includegraphics[scale=0.5]{Two_way_plots/Twoway_a1leta5l2_m020304.pdf}\caption{Effective potential (left) and time evolution (right) of scalar perturbations, with varying $m$, of the two-way wormhole with  $\ell = 2$, $\ell_\eta = 5$ and $\alpha = 1$. As can be seen, increasing the mass affects the potential in a similar manner with increasing the throat $\alpha$, in the sense that the potential peak lessens. However, the field exhibits a substantially different behavior. We observe the echoes arriving in smaller time scales for heavier compact objects despite the fact that the reflective AdS boundary is not affected by the different values of mass.\textcolor{red}{I am not sure how is this behavior explained...It seems very similar with the Rinaldi(see Figures 4,5 in our \href{https://arxiv.org/pdf/2109.02678.pdf}{previous paper}) case but there we had a horizon that was moving outwards due to the increase of the mass...Here we do not a have a horizon. } For high enough masses, a new potential well forms at the wormhole throat. In this case, echoes may arise due to trapping modes from a primary region between the PS and the asymptotic AdS boundary, and a secondary region located at the wormhole throat.f} \label{fig:4}\end{figure}
%------------------------------------------
\subsection{One-way Wormhole}
\begin{figure}[]
    \centering
    \includegraphics[scale=0.479]{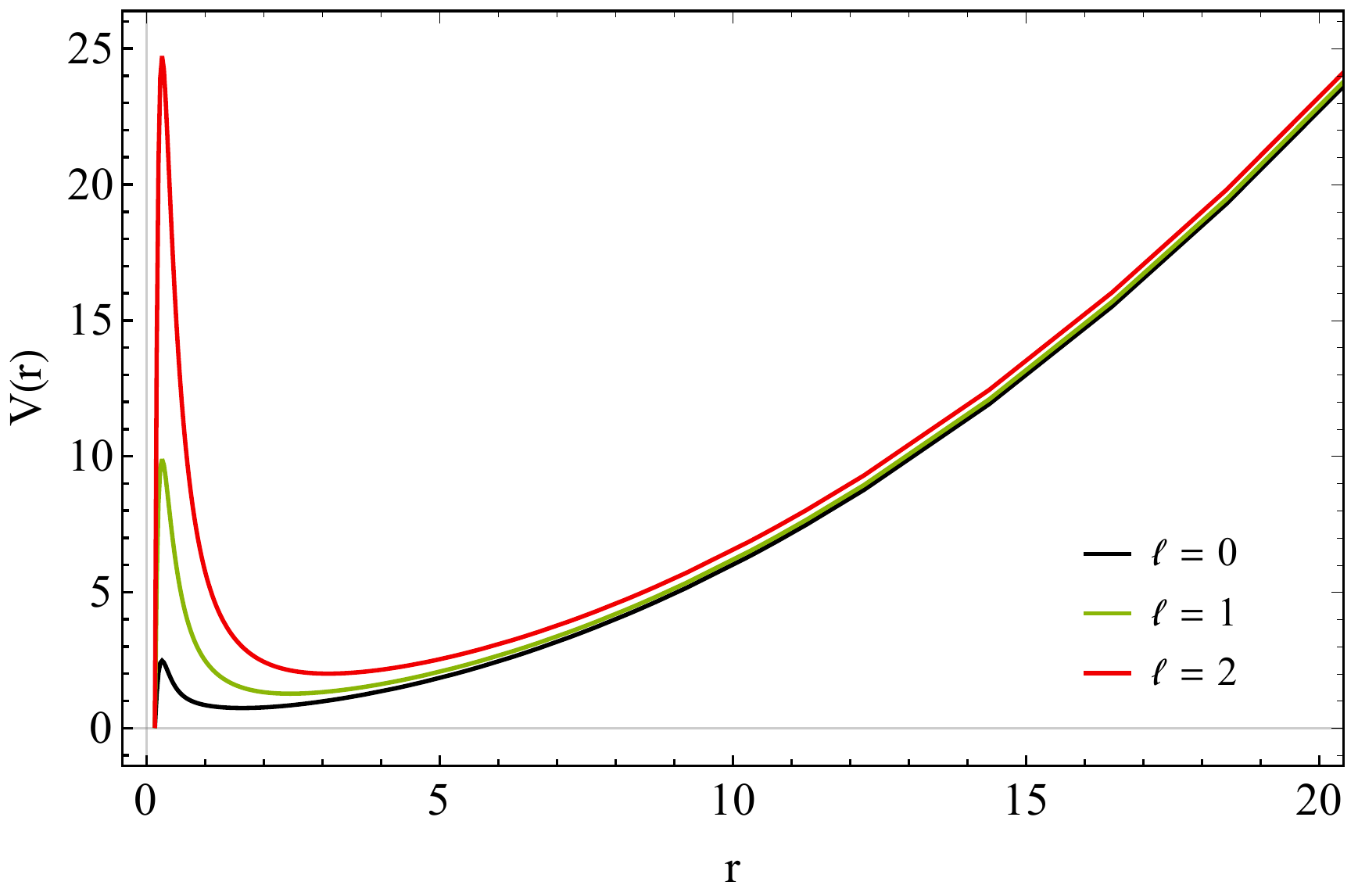}
    \hspace{0.2 cm}
    \includegraphics[scale=0.5]{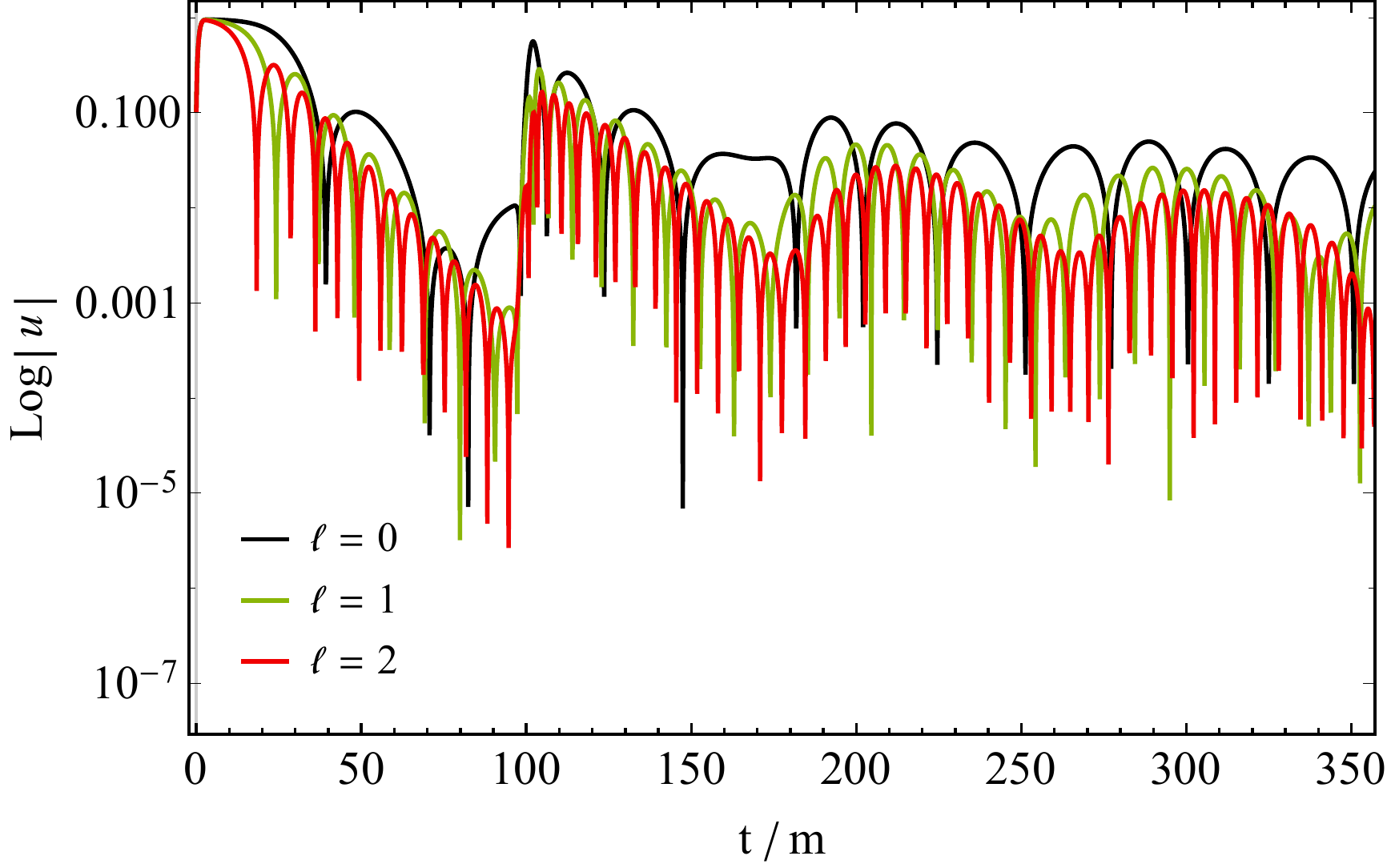}
    \caption{Effective potential (left) and time evolution (right) of scalar perturbations, with varying $\ell$, of the one-way wormhole with  $m = 0.1$ and $\ell_\eta = 1$.
    Parameter $\ell$ has the usual effect. Increases the potential barrier which leads to more energy being extracted from the PS, leading to signal with higher frequency.}
    \label{fig:7}
\end{figure}

%------------------------------------------
In what follows, we set $\alpha = r_h$ in order for our compact object \eqref{3.9} to describe a one-way wormhole with a null throat. Fig.~\ref{fig:7}  displays the evolution of a linear scalar perturbation field on such a  background. The temporal response exhibits echoes, as in the two-way wormhole case above, which follow the initial ringdown. In a similar manner, the $\ell=0$ perturbations do not significantly excite the PS of the compact object, thus the echoes are not as oscillatory as the ones obtained for $\ell>0$.

\begin{figure}[]
    \centering
    \includegraphics[scale=0.479]{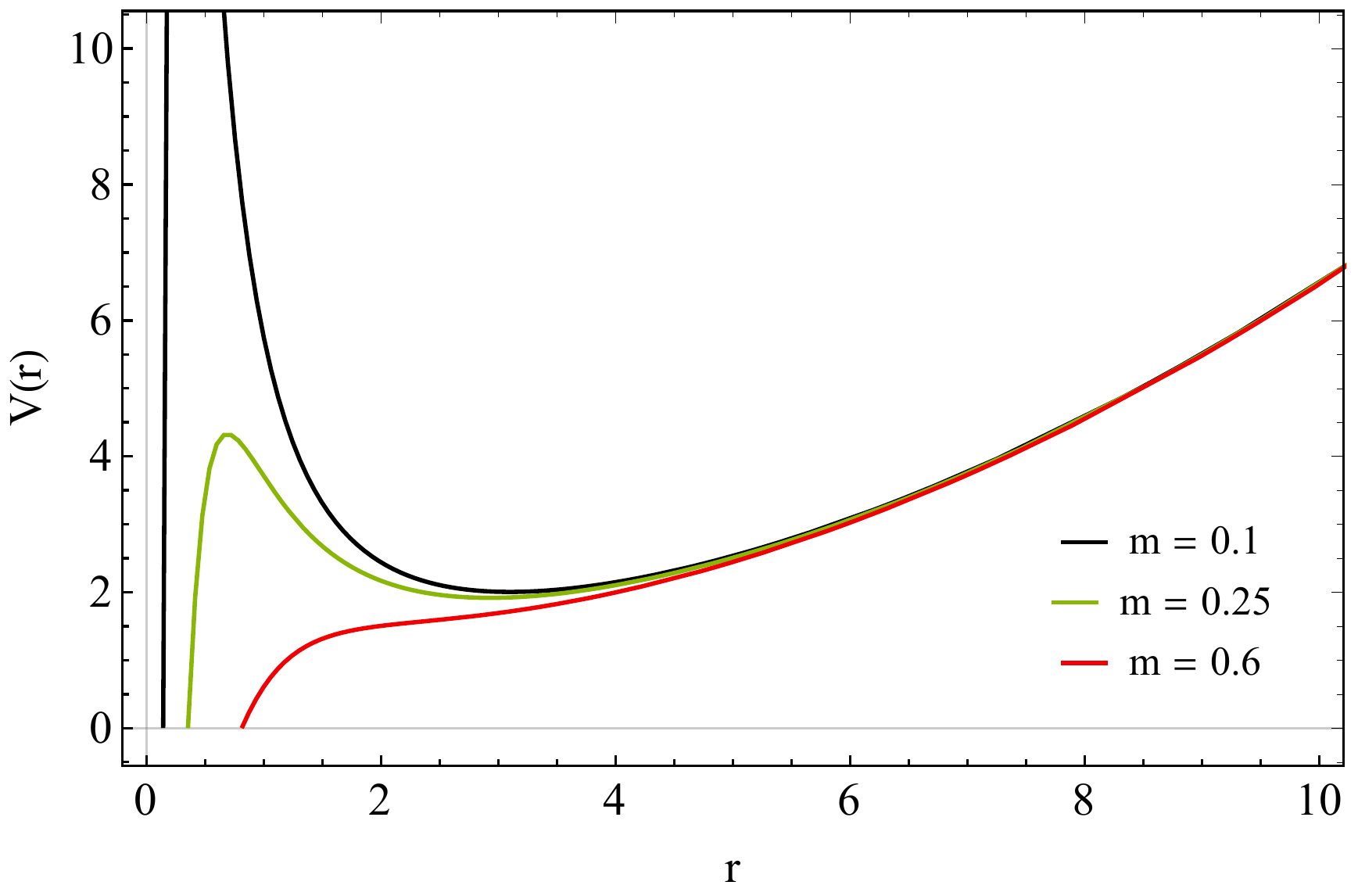}
    \hspace{0.2 cm}
    \includegraphics[scale=0.5]{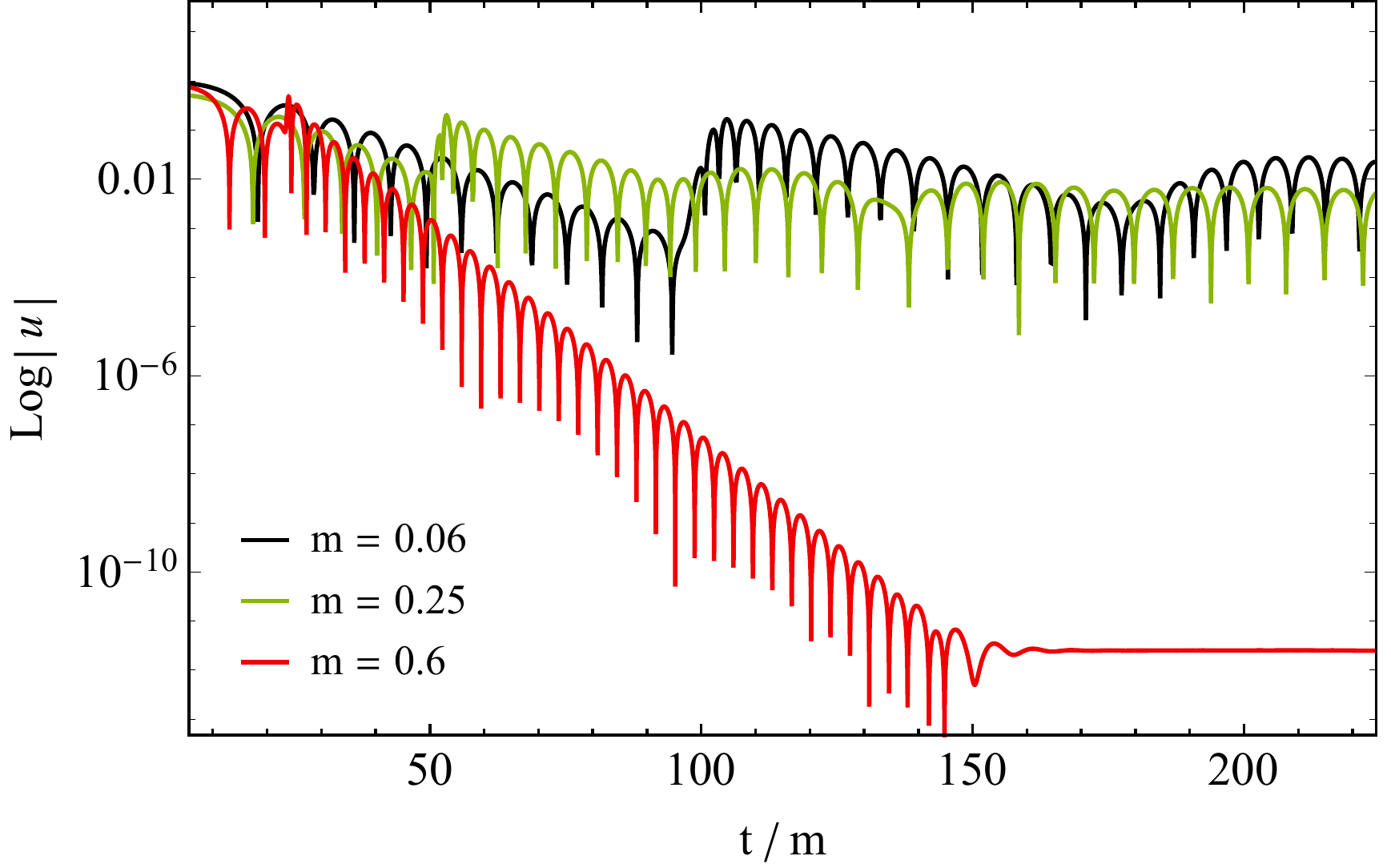}
    \caption{Effective potential (left) and time evolution (right) of scalar perturbations, with varying $m$, of the one-way wormhole with  $\ell = 2$ and $\ell_\eta = 1$.}
    \label{fig:8}
\end{figure}

\begin{figure}[]
    \centering
    \includegraphics[scale=0.479]{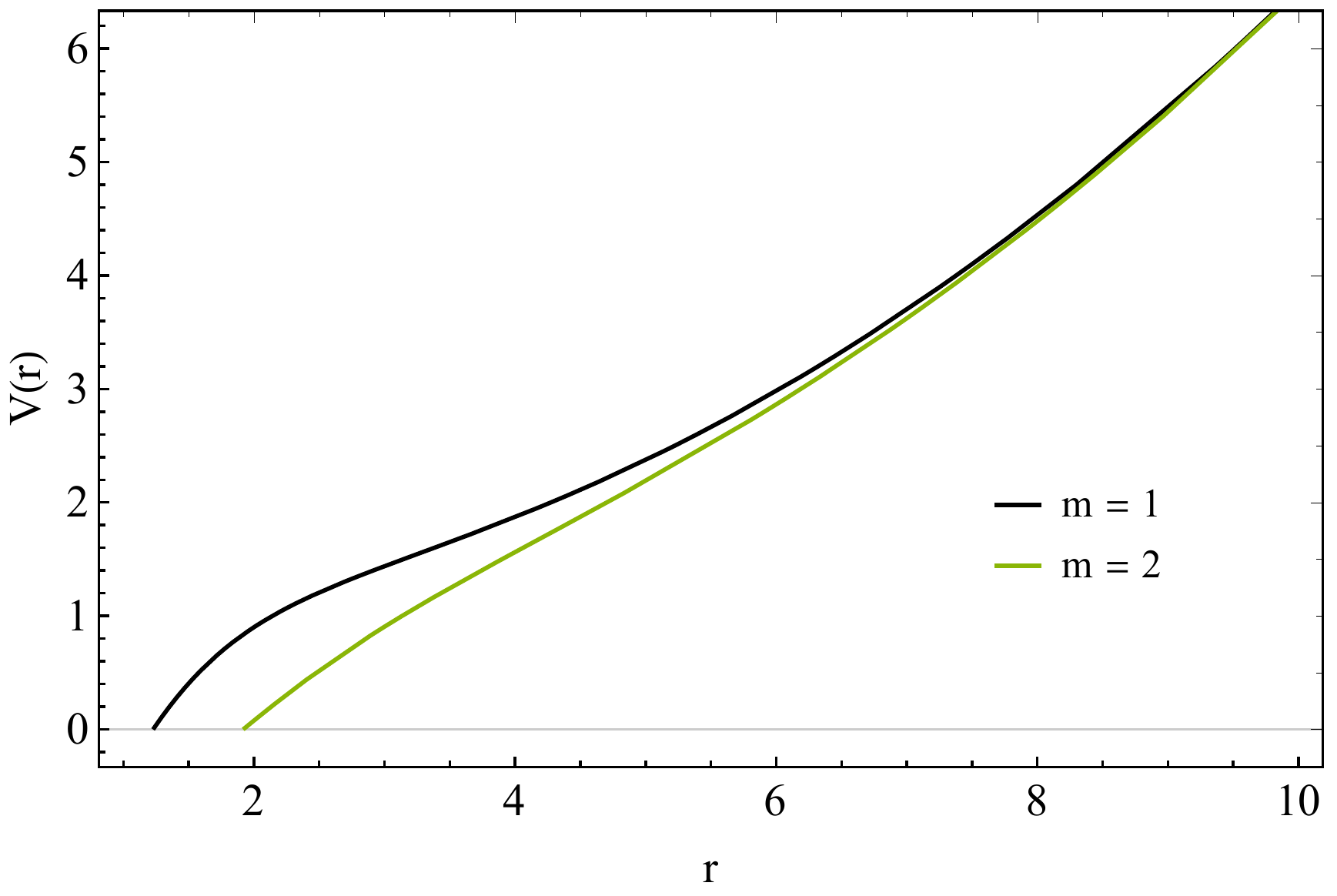}
    \hspace{0.2 cm}
    \includegraphics[scale=0.5]{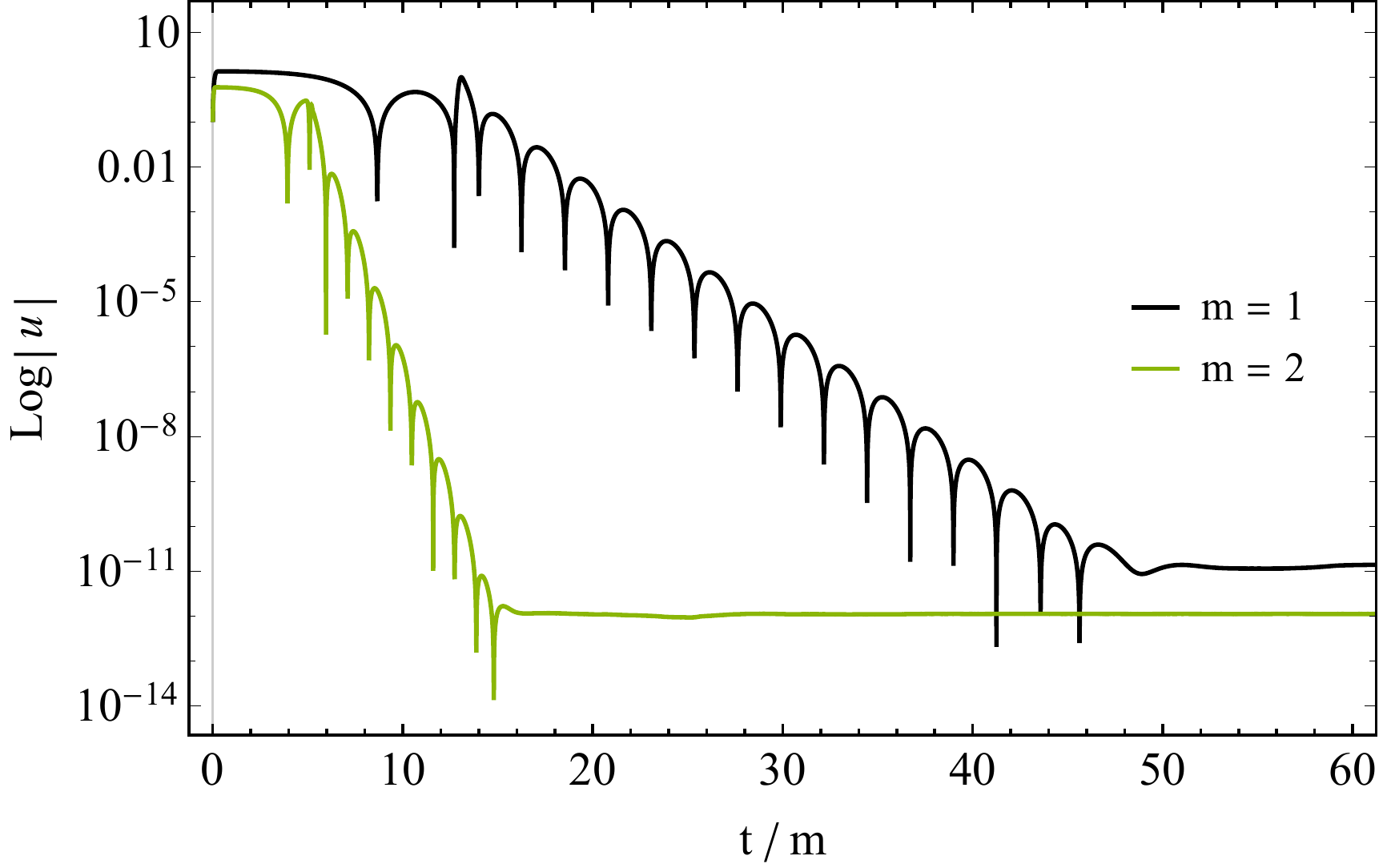}
    \caption{Effective potential (left) and time evolution (right) of scalar perturbations, with varying $m$, of the one-way wormhole with  $\ell = 2$ and $\ell_\eta = 1$.}
    \label{fig:9}
\end{figure}

In Fig.~\ref{fig:8} the effect of mass on the temporal response is illustrated. As $m$ grows the echoes are replaced by quasinormal oscillations, while any further increment leads to a single quasinormal ringdown followed by a late-time tail. We conclude that this behavior stems from the shape of the effective potential which decreases in amplitude as $m$ increases. This leads to an increasingly smaller region where trapped modes can occur, and thus the quasinormal ringing dominates over the echoes which are suppressed. Any further increment of $m$ results in the decreasing of the damping time of the quasinormal ringing (see Fig. \ref{fig:9}). Moreover, note that the field settles down to a constant non zero value after the initial rigndown. Such a behavior is also observed \cite{Chatzifotis:2021pak} when axial perturbations are present in the  black hole \cite{Rinaldi:2012vy} spacetime. Moreover, in contrast to the previous case, the echoes found here have significantly smaller amplitudes when compared to the initial ringdown. This pattern emerges because of the presence of an event horizon at the would-be wormhole throat of our solution, which changes drastically the boundary conditions of our problem and introduces  energy dissipation. Hence, such compact objects respond to perturbations in a similar manner to BHs, since any extra information indicating the presence of a wormhole is locked behind the event horizon.

%The decay rate of perturbations follows an exponential fall-off which is more evident for the spherically symmetric $\ell=0$ case. This behavior is in contrast to that of asymptotically flat BH perturbations, where the quasinormal ringing gives way to a power-law cutoff \cite{Gundlach:1993tp, Leaver:1986gd, Gundlach:1993tn}, and its occurrence is related reflective nature of the AdS boundary.
%---------------------------------------------
%---------------------------------------------
%\begin{figure}[H]\centering\includegraphics[scale=0.479]{One_way_plots/Oneway_leta1m025_l012Veff.pdf}\hspace{0.2 cm}\includegraphics[scale=0.5]{One_way_plots/Oneway_leta1m025_l012.pdf}\caption{Effective potential (left) and time evolution (right) of scalar perturbations, with varying $\ell$, $m = 0.25$ and $\ell_\eta = 1$. }\label{fig:8}\end{figure}
%---------------------------------------------
%---------------------------------------------

\section{Conclusions }
\label{sec6}

We studied the deformation of  a black hole to a wormhole using disformal transformations. We  considered the metric described by a hairy black hole solution of a subclass of the Horndeski theory in which the scalar field is coupled kinetically to Einstein tensor  as a seed physical  metric. Then, by applying a disformal transformation, we get a disformed metric  describing a wormhole geometry. This metric interpolates between black hole, regular black hole, one-way wormhole and two-way wormhole solution depending on a critical parameter $a$ which grows  from zero to values larger than the black hole event horizon, describing the wormhole throat. Calculating the energy-momentum tensor of the wormhole geometry, we find that it is a  deformation of  the energy-momentum tensor of the original seed physical  metric we considered.

We tested the possible violations of the null energy conditions as the $a$ parameter is increasing and the solutions shifts from the original black hole to the  wormhole. Fixing  the event horizon of the original black hole to be the null hypersurface $r=1$, we find that in the case that $a$ is non-zero as the  throat radius grows the violation of the null energy condition is stronger. We argue that the violation of the null energy condition is due to the presence in the energy-momentum tensor  of the term that deforms of  the energy-momentum tensor of the original seed black hole solution.

Using the time evolution of scalar perturbations in our shifting geometry, we studied the temporal response of linear massless scalar field perturbations on the discussed  wormhole solutions. Our investigation indicates that similar effects arise in the late-time ringdown for both compact objects. After the initial ringdown, the test field response exhibits
echoes, with timescales proportional to the non-minimal coupling constant $\ell_\eta$. The beating pattern is similar to echoes from quantum corrected BHs \cite{Chatzifotis:2020oqr} even though, the compact objects considered here do not contain any quantum corrections. The effective AdS asymptotics of both solutions, force the partially reflected waves from the PS to mirror off the AdS boundary and re-perturb the PS to give rise to the observed echoes.

A main difference between the two compact objects is the existence of a horizon on the one-way wormhole which introduces energy dissipation resulting in echoes whose amplitude decrease with time, in contrast to the two-way wormhole setup. Mathematically this is understood by the drastic change in the boundary conditions imposed on the scattering problem. Regarding the two-way wormhole, the constancy of the amplitude of echoes may be an indication of the existence of normal oscillation modes, as well as potential instabilities, similar to that found in \cite{Destounis:2019hca}.

We expect that the gravitational perturbations
of the  metric would give us decisive  information on the stability of the geometry we found, which we leave it for feature work. It would also be interesting to extend this study in the case that the physical seed metric is the one discussed in \cite{Babichev:2013cya} which is also a hairy black hole solution of the same subclass of the Horndeski theory with a scalar field kinetically coupled to Einstein tensor. The new feature of this solution is that the scalar field is time dependent and it would be interesting to study its effect on the disformal deformation of this seed metric.

%---------------------------------------------------

\appendix
\section{Appendix}

In this Appendix we present the transformation of the various terms in the gravitational tensors under the disformal transformation under consideration.
The inverse of the disformed metric can be readily given by the Sherman-Morrison formula as
\begin{equation}
	\label{A.1}
	\hat{g}^{\mu\nu}=\frac{1}{\Omega^2}\left(g^{\mu\nu}-\frac{W}{\Omega^2+W X}\partial^\mu\Phi\partial^\nu\Phi\right)~,
\end{equation}
where the indices of $\partial^\mu\Phi\partial^\nu\Phi$ were raised using the original seed metric, $g^{\mu\nu}$. Formula (\Ref{A.1}) yields the obvious requirement that
\begin{align}
	\label{A.2}
	&\Omega\neq0~,\\
	\label{A.3}
	&\Omega^2+W X\neq0~.
\end{align}
In addition, in order to satisfy the Lorentzian signature, the following conditions must hold \cite{Bettoni:2013diz}
\begin{align}
	\label{A.4}
	&\Omega^2>0~,\\
	\label{A.5}
	&\Omega^2+W X>0~.
\end{align}
The use of (\Ref{A.1}) gives us the transformed Levi Civita connection as follows
\begin{eqnarray*}
	\hat{\Gamma}^{\alpha}_{\,\,\,\mu\nu}=
	&&\frac{1}{2}\left[\frac{1}{\Omega^2}\left(g^{\alpha\beta}-\frac{W}{\Omega^2+W X}\partial^\alpha\Phi\partial^\beta\Phi\right)\right]\times \\
	&&\left[(\partial_\nu \Omega^2) g_{\beta\mu}+(\partial_\mu \Omega^2) g_{\nu\beta}-(\partial_\beta \Omega^2) g_{\mu\nu}\right.\\
	&&\left.+\Omega^2\left(\partial_\nu g_{\beta\mu}+\partial_\mu g_{\nu\beta}-\partial_\beta g_{\mu\nu}\right)\right.\\
	&&\left.+\partial_\nu W\partial_\beta \Phi\partial_\mu \Phi+\partial_\mu W\partial_\nu \Phi\partial_\beta \Phi-\partial_\beta W\partial_\mu \Phi\partial_\nu \Phi\right.\\
	&&\left.+W\left(\partial_\nu(\partial_\beta\Phi)\partial_\mu\Phi+\partial_\beta\Phi\partial_\nu(\partial_\mu\Phi)\right)\right.\\
	&&\left.+W\left(\partial_\mu(\partial_\nu\Phi)\partial_\beta\Phi+\partial_\nu\Phi\partial_\mu(\partial_\beta\Phi)\right)\right.\\
	&&\left.-W\left(\partial_\beta(\partial_\mu\Phi)\partial_\nu\Phi-\partial_\mu\Phi\partial_\beta(\partial_\nu\Phi)\right)\right]~.
\end{eqnarray*}
For simplification, we set the conformal factor $\Omega=1$. This greatly simplifies our calculations, since the differentiation of the conformal factos vanishes. The result for the connection is found to be
\begin{equation}
	\label{A.7}
	\hat{\Gamma}^{\alpha}_{\,\,\,\mu\nu}=\Gamma^{\alpha}_{\,\,\,\mu\nu}+D^{\alpha}_{\,\,\,\mu\nu}~,
\end{equation}
where $D^{\alpha}_{\,\,\,\mu\nu}$ is the disformal term in the connection and reads
\begin{equation}
	\label{A.8}
	D^{\alpha}_{\,\,\,\mu\nu}=\frac{1}{2}\hat{g}^{\alpha\beta}(\partial_\nu W\partial_\beta \Phi\partial_\mu \Phi+\partial_\mu W\partial_\nu \Phi\partial_\beta \Phi-\partial_\beta W\partial_\mu \Phi\partial_\nu \Phi)+\nabla_\nu(\partial_\mu\Phi)\partial^{\alpha}\Phi\frac{W}{1+WX}~.	
\end{equation}
Note that the metric term in (\Ref{A.8}) is the disformed one, which under the constraint of $\Omega=1$, readily gives us that $\displaystyle \hat{g}^{\mu\nu}=\left(g^{\mu\nu}-\frac{W}{1+W X}\partial^\mu\Phi\partial^\nu\Phi\right)$. A second important note here is that $D^{\alpha}_{\,\,\,\mu\nu}$ retains the symmetry of the Levi Civita connection in its bottom two indices and is a tensor, as expected from the difference of the two connections. The simplicity of (\Ref{A.7}) means that we can easily find the rest of the gravitational tensors as follows
\begin{align*}
	\hat{R}^{\rho}_{\,\,\,\sigma\mu\nu}&=\partial_{\mu} \hat{\Gamma}^{\rho}_{\,\,\,\sigma\nu}-\partial_{\nu} \hat{\Gamma}^{\rho}_{\,\,\,\sigma\mu}+\hat{\Gamma}^{\rho}_{\,\,\,\kappa\mu}\hat{\Gamma}^{\kappa}_{\,\,\,\sigma\nu}-\hat{\Gamma}^{\rho}_{\,\,\,\kappa\nu}\hat{\Gamma}^{\kappa}_{\,\,\,\sigma\mu}\\
	&=\partial_{\mu} \Gamma^{\rho}_{\,\,\,\sigma\nu}+\partial_{\mu} D^{\rho}_{\,\,\,\sigma\nu}-\partial_{\nu}\Gamma^{\rho}_{\,\,\,\sigma\mu}-\partial_{\nu} D^{\rho}_{\,\,\,\sigma\mu}
	+\left(\Gamma^{\rho}_{\,\,\,\kappa\mu}+D^{\rho}_{\,\,\,\kappa\mu}\right)\left(\Gamma^{\kappa}_{\,\,\,\sigma\nu}+D^{\kappa}_{\,\,\,\sigma\nu}\right)
	-\left(\Gamma^{\rho}_{\,\,\,\kappa\nu}+D^{\rho}_{\,\,\,\kappa\nu}\right)\left(\Gamma^{\kappa}_{\,\,\,\sigma\mu}+D^{\kappa}_{\,\,\,\sigma\mu}\right)\\
	&=	\partial_{\mu} \Gamma^{\rho}_{\,\,\,\sigma\nu}-\partial_{\nu} \Gamma^{\rho}_{\,\,\,\sigma\mu}+\Gamma^{\rho}_{\,\,\,\kappa\mu}\Gamma^{\kappa}_{\,\,\,\sigma\nu}-\Gamma^{\rho}_{\,\,\,\kappa\nu}\Gamma^{\kappa}_{\,\,\,\sigma\mu}\\
	&+\partial_{\mu} D^{\rho}_{\,\,\,\sigma\nu}+\Gamma^{\rho}_{\,\,\,\kappa\mu}D^{\kappa}_{\,\,\,\sigma\nu}
-\Gamma^{\kappa}_{\,\,\,\sigma\mu}D^{\rho}_{\,\,\,\kappa\nu}-\partial_{\nu} D^{\rho}_{\,\,\,\sigma\mu}-\Gamma^{\rho}_{\,\,\,\kappa\nu}D^{\kappa}_{\,\,\,\sigma\mu}
+\Gamma^{\kappa}_{\,\,\,\sigma\nu}D^{\rho}_{\,\,\,\kappa\mu}
+D^{\rho}_{\,\,\,\kappa\mu}D^{\kappa}_{\,\,\,\sigma\nu}-D^{\rho}_{\,\,\,\kappa\nu}D^{\kappa}_{\,\,\,\sigma\mu}~,
\end{align*}
from which we get
\begin{equation}
	\label{A.9}
	\rightarrow	\hat{R}^{\rho}_{\,\,\,\sigma\mu\nu}	=R^{\rho}_{\,\,\,\sigma\mu\nu}+\nabla_\mu D^{\rho}_{\,\,\,\sigma\nu}-\nabla_\nu D^{\rho}_{\,\,\,\sigma\mu}+D^{\rho}_{\,\,\,\kappa\mu}D^{\kappa}_{\,\,\,\sigma\nu}-D^{\rho}_{\,\,\,\kappa\nu}D^{\kappa}_{\,\,\,\sigma\mu}~.
\end{equation}
Using the fact that $\delta^{\alpha}_{\beta}=g^{\alpha\rho}g_{\beta\rho}$ is invariant under disformal transformations, we find that the disformed Ricci tensor reads
\begin{equation}
	\label{A.10}
	\rightarrow	\hat{R}_{\sigma\nu}	=R_{\sigma\nu}+\nabla_\mu D^{\mu}_{\,\,\,\sigma\nu}-\nabla_\nu D^{\mu}_{\,\,\,\sigma\mu}+D^{\mu}_{\,\,\,\kappa\mu}D^{\kappa}_{\,\,\,\sigma\nu}-D^{\mu}_{\,\,\,\kappa\nu}D^{\kappa}_{\,\,\,\sigma\mu}~.
\end{equation}
Finally, contracting (\Ref{A.10}) with the disformed metric we have the following result for the Ricci scalar
\begin{equation}
	\label{A.11}
	\hat{R}=R+\nabla_\mu D^{\mu\nu}_{\,\,\,\,\,\,\nu}-\nabla_\nu D^{\mu\nu}_{\,\,\,\,\,\,\mu}+D^{\mu}_{\,\,\,\kappa\mu}D^{\kappa\nu}_{\,\,\,\,\,\,\nu}-D^{\mu}_{\,\,\,\kappa\nu}D^{\kappa\nu}_{\,\,\,\,\,\,\mu}-\frac{W\partial^{\sigma}\Phi\partial^{\nu}\Phi}{1+WX}(R_{\sigma\nu}+\nabla_\mu D^{\mu}_{\,\,\,\sigma\nu}-\nabla_\nu D^{\mu}_{\,\,\,\sigma\mu}+D^{\mu}_{\,\,\,\kappa\mu}D^{\kappa}_{\,\,\,\sigma\nu}-D^{\mu}_{\,\,\,\kappa\nu}D^{\kappa}_{\,\,\,\sigma\mu}).
\end{equation}
As such the Einstein tensor reads
\begin{align*}
	\hat{G}_{\mu\nu}&=\hat{R}_{\mu\nu}-\frac{1}{2}\hat{g}_{\mu\nu}\hat{R}\\
	&=R_{\mu\nu}+\nabla_\kappa D^{\kappa}_{\,\,\,\mu\nu}-\nabla_\nu D^{\kappa}_{\,\,\,\mu\kappa}+D^{\kappa}_{\,\,\,\lambda\kappa}D^{\lambda}_{\,\,\,\mu\nu}-D^{\kappa}_{\,\,\,\lambda\nu}D^{\lambda}_{\,\,\,\mu\kappa}\\
	&-\frac{1}{2}(g_{\mu\nu}+W\partial_\mu\Phi\partial_\nu\Phi)\Big[R+\nabla_\kappa D^{\kappa\lambda}_{\,\,\,\,\,\,\lambda}-\nabla_\lambda D^{\kappa\lambda}_{\,\,\,\,\,\,\kappa}+D^{\kappa}_{\,\,\,\sigma\kappa}D^{\sigma\lambda}_{\,\,\,\,\,\,\lambda}-D^{\kappa}_{\,\,\,\sigma\lambda}D^{\sigma\lambda}_{\,\,\,\,\,\,\kappa}\\
	&\qquad\qquad\qquad\qquad\qquad
	-\frac{W\partial^{\sigma}\Phi\partial^{\lambda}\Phi}{1+WX}(R_{\sigma\lambda}+\nabla_\kappa D^{\kappa}_{\,\,\,\sigma\lambda}-\nabla_\lambda D^{\kappa}_{\,\,\,\sigma\kappa}+D^{\kappa}_{\,\,\,\rho\kappa}D^{\rho}_{\,\,\,\sigma\lambda}
-D^{\kappa}_{\,\,\,\rho\lambda}D^{\rho}_{\,\,\,\sigma\kappa})\Big]~,
\end{align*}
which finally yields that
\begin{equation}
	\label{A.12}
	\hat{G}_{\mu\nu}=G_{\mu\nu}+S_{\mu\nu}~,
\end{equation}
where
\begin{align}
	\label{A.13}
	\nonumber
	S_{\mu\nu}=&\nabla_\kappa D^{\kappa}_{\,\,\,\mu\nu}-\nabla_\nu D^{\kappa}_{\,\,\,\mu\kappa}+D^{\kappa}_{\,\,\,\lambda\kappa}D^{\lambda}_{\,\,\,\mu\nu}-D^{\kappa}_{\,\,\,\lambda\nu}D^{\lambda}_{\,\,\,\mu\kappa}-\frac{1}{2}W\partial_\mu\Phi\partial_\nu\Phi R\\
	\nonumber
	&-\frac{1}{2}\hat{g}_{\mu\nu}\Big[\nabla_\kappa D^{\kappa\lambda}_{\,\,\,\,\,\,\lambda}-\nabla_\lambda D^{\kappa\lambda}_{\,\,\,\,\,\,\kappa}+D^{\kappa}_{\,\,\,\sigma\kappa}D^{\sigma\lambda}_{\,\,\,\,\,\,\lambda}-D^{\kappa}_{\,\,\,\sigma\lambda}D^{\sigma\lambda}_{\,\,\,\,\,\,\kappa}\\
	&\qquad\qquad-\frac{W\partial^{\sigma}\Phi\partial^{\lambda}\Phi}{1+WX}(R_{\sigma\lambda}+\nabla_\kappa D^{\kappa}_{\,\,\,\sigma\lambda}-\nabla_\lambda D^{\kappa}_{\,\,\,\sigma\kappa}+D^{\kappa}_{\,\,\,\rho\kappa}D^{\rho}_{\,\,\,\sigma\lambda}-D^{\kappa}_{\,\,\,\rho\lambda}D^{\rho}_{\,\,\,\sigma\kappa})\Big]~.
\end{align}


\begin{thebibliography}{99}

%\cite{Bekenstein:1992pj}
\bibitem{Bekenstein:1992pj}
J.~D.~Bekenstein,
``The Relation between physical and gravitational geometry,''
Phys. Rev. D \textbf{48}, 3641-3647 (1993)
%doi:10.1103/PhysRevD.48.3641
[arXiv:gr-qc/9211017 [gr-qc]].
%359 citations counted in INSPIRE as of 07 Apr 2021

\bibitem{Horndeski:1974wa}
	G.W. Horndeski, {Second-order scalar-tensor field equations in a
		four-dimensional space}. Int. J. Theor. Phys. \textbf{10}, 363 (1974).

%\cite{Bettoni:2013diz}
\bibitem{Bettoni:2013diz}
D.~Bettoni and S.~Liberati,
`Disformal invariance of second order scalar-tensor theories: Framing the Horndeski action,''
Phys. Rev. D \textbf{88}, 084020 (2013)
%doi:10.1103/PhysRevD.88.084020
[arXiv:1306.6724 [gr-qc]].
%192 citations counted in INSPIRE as of 10 Sep 2021

%\cite{Achour:2016rkg}
\bibitem{Achour:2016rkg}
J.~Ben Achour, D.~Langlois and K.~Noui,
``Degenerate higher order scalar-tensor theories beyond Horndeski and disformal transformations,''
Phys. Rev. D \textbf{93}, no.12, 124005 (2016)
%doi:10.1103/PhysRevD.93.124005
[arXiv:1602.08398 [gr-qc]].
%191 citations counted in INSPIRE as of 08 Apr 2021


%\cite{Deffayet:2020ypa}
\bibitem{Deffayet:2020ypa}
C.~Deffayet and S.~Garcia-Saenz,
``Degeneracy, matter coupling, and disformal transformations in scalar-tensor theories,''
Phys. Rev. D \textbf{102}, no.6, 064037 (2020)
%doi:10.1103/PhysRevD.102.064037
[arXiv:2004.11619 [hep-th]].
%1 citations counted in INSPIRE as of 08 Apr 2021

%\cite{Tsujikawa:2014uza}
\bibitem{Tsujikawa:2014uza}
S.~Tsujikawa,
``Disformal invariance of cosmological perturbations in a generalized class of Horndeski theories,''
JCAP \textbf{04}, 043 (2015)
%doi:10.1088/1475-7516/2015/04/043
[arXiv:1412.6210 [hep-th]].
%44 citations counted in INSPIRE as of 08 Apr 2021

%\cite{Domenech:2015hka}
\bibitem{Domenech:2015hka}
G.~Dom\`enech, A.~Naruko and M.~Sasaki,
``Cosmological disformal invariance,''
JCAP \textbf{10}, 067 (2015)
%doi:10.1088/1475-7516/2015/10/067
[arXiv:1505.00174 [gr-qc]].
%78 citations counted in INSPIRE as of 20 Nov 2021



%\cite{Minamitsuji:2016hkk}
\bibitem{Minamitsuji:2016hkk}
M.~Minamitsuji and H.~O.~Silva,
``Relativistic stars in scalar-tensor theories with disformal coupling,''
Phys. Rev. D \textbf{93}, no.12, 124041 (2016)
%doi:10.1103/PhysRevD.93.124041
[arXiv:1604.07742 [gr-qc]].
%32 citations counted in INSPIRE as of 08 Apr 2021



%\cite{Langlois:2017mdk}
\bibitem{Langlois:2017mdk}
D.~Langlois,
``Degenerate Higher-Order Scalar-Tensor (DHOST) theories,''
[arXiv:1707.03625 [gr-qc]].
%25 citations counted in INSPIRE as of 11 Oct 2021

%\cite{Langlois:2018dxi}
\bibitem{Langlois:2018dxi}
D.~Langlois,
``Dark energy and modified gravity in degenerate higher-order scalar\textendash{}tensor (DHOST) theories: A review,''
Int. J. Mod. Phys. D \textbf{28} (2019) no.05, 1942006
%doi:10.1142/S0218271819420069
[arXiv:1811.06271 [gr-qc]].
%97 citations counted in INSPIRE as of 11 Oct 2021


%\cite{BenAchour:2020wiw}
\bibitem{BenAchour:2020wiw}
J.~Ben Achour, H.~Liu and S.~Mukohyama,
``Hairy black holes in DHOST theories: Exploring disformal transformation as a solution-generating method,''
JCAP \textbf{02} (2020), 023
%doi:10.1088/1475-7516/2020/02/023
[arXiv:1910.11017 [gr-qc]].
%28 citations counted in INSPIRE as of 11 Oct 2021



%\cite{Faraoni:2021gdl}
\bibitem{Faraoni:2021gdl}
V.~Faraoni and A.~Leblanc,
``Disformal mappings of spherical DHOST geometries,''
[arXiv:2107.03456 [gr-qc]].
%0 citations counted in INSPIRE as of 19 Jul 2021

%\cite{Anson:2020trg}
\bibitem{Anson:2020trg}
T.~Anson, E.~Babichev, C.~Charmousis and M.~Hassaine,
``Disforming the Kerr metric,''
JHEP \textbf{01}, 018 (2021)
%doi:10.1007/JHEP01(2021)018
[arXiv:2006.06461 [gr-qc]].
%21 citations counted in INSPIRE as of 24 Nov 2021

%\cite{BenAchour:2020fgy}
\bibitem{BenAchour:2020fgy}
J.~Ben Achour, H.~Liu, H.~Motohashi, S.~Mukohyama and K.~Noui,
``On rotating black holes in DHOST theories,''
JCAP \textbf{11}, 001 (2020)
%doi:10.1088/1475-7516/2020/11/001
[arXiv:2006.07245 [gr-qc]].
%22 citations counted in INSPIRE as of 20 Nov 2021


%\cite{Erices:2021uyu}
\bibitem{Erices:2021uyu}
C.~Erices, P.~Filis and E.~Papantonopoulos,
``Hairy black holes in disformal scalar-tensor gravity theories,''
Phys. Rev. D \textbf{104} (2021) no.2, 024031
%doi:10.1103/PhysRevD.104.024031
[arXiv:2104.05644 [gr-qc]].
%3 citations counted in INSPIRE as of 13 Oct 2021

%\cite{Rinaldi:2012vy}
	\bibitem{Rinaldi:2012vy}
	M.~Rinaldi,
	``Black holes with non-minimal derivative coupling,''
	Phys.\ Rev.\ D {\bf 86}, 084048 (2012)
	% doi:10.1103/PhysRevD.86.084048
	[arXiv:1208.0103 [gr-qc]].
	%%CITATION = doi:10.1103/PhysRevD.86.084048;%%
	%154 citations counted in INSPIRE as of 10 Jan 2020

%\cite{Vlachos:2021weq}
	\bibitem{Vlachos:2021weq}
	C.~Vlachos, E.~Papantonopoulos and K.~Destounis,
	``Echoes of Compact Objects in Scalar-Tensor Theories of Gravity,''
	Phys. Rev. D \textbf{103}, no.4, 044042 (2021)
	%doi:10.1103/PhysRevD.103.044042
	[arXiv:2101.12196 [gr-qc]].
	%0 citations counted in INSPIRE as of 17 Apr 2021

%\cite{Chatzifotis:2021pak}
	\bibitem{Chatzifotis:2021pak}
	N.~Chatzifotis, C.~Vlachos, K.~Destounis and E.~Papantonopoulos,
	``Stability of black holes with non-minimally coupled scalar hair,''
	[arXiv:2109.02678 [gr-qc]].
	%0 citations counted in INSPIRE as of 24 Sep 2021

\bibitem{Hochberg:1997wp}
D.~Hochberg and M.~Visser,
``Geometric structure of the generic static traversable wormhole throat,''
Phys. Rev. D \textbf{56}, 4745-4755 (1997)
%doi:10.1103/PhysRevD.56.4745
[arXiv:gr-qc/9704082 [gr-qc]].
%238 citations counted in INSPIRE as of 19 Oct 2021

%\cite{Simpson:2018tsi}
	\bibitem{Simpson:2018tsi}
	A.~Simpson and M.~Visser,
	``Black-bounce to traversable wormhole,''
	JCAP \textbf{02}, 042 (2019)
	%doi:10.1088/1475-7516/2019/02/042
	[arXiv:1812.07114 [gr-qc]].
	%48 citations counted in INSPIRE as of 30 Aug 2021

%\cite{Stuchlik:2021tcn}
\bibitem{Stuchlik:2021tcn}
Z.~Stuchl\'\i{}k and J.~Vrba,
``Epicyclic Oscillations around Simpson\textendash{}Visser Regular Black Holes and Wormholes,''
Universe \textbf{7}, no.8, 279 (2021)
%doi:10.3390/universe7080279
[arXiv:2108.09562 [gr-qc]].
%0 citations counted in INSPIRE as of 30 Aug 2021

%\cite{Chakrabarti:2021gqa}
\bibitem{Chakrabarti:2021gqa}
S.~Chakrabarti and S.~Kar,
``Wormhole geometry from gravitational collapse,''
Phys. Rev. D \textbf{104}, no.2, 024071 (2021)
%doi:10.1103/PhysRevD.104.024071
[arXiv:2106.14761 [gr-qc]].
%0 citations counted in INSPIRE as of 30 Aug 2021

%\cite{Domenech:2019syf}
\bibitem{Domenech:2019syf}
G.~Dom\`enech, A.~Naruko, M.~Sasaki and C.~Wetterich,
``Could the black hole singularity be a field singularity?,''
Int. J. Mod. Phys. D \textbf{29}, no.03, 2050026 (2020)
%doi:10.1142/S0218271820500261
[arXiv:1912.02845 [gr-qc]].
%11 citations counted in INSPIRE as of 20 Nov 2021


%\cite{Franzin:2021vnj}
\bibitem{Franzin:2021vnj}
E.~Franzin, S.~Liberati, J.~Mazza, A.~Simpson and M.~Visser,
``Charged black-bounce spacetimes,''
JCAP \textbf{07}, 036 (2021)
%doi:10.1088/1475-7516/2021/07/036
[arXiv:2104.11376 [gr-qc]].
%4 citations counted in INSPIRE as of 30 Aug 2021

%\cite{Islam:2021ful}
\bibitem{Islam:2021ful}
S.~U.~Islam, J.~Kumar and S.~G.~Ghosh,
``Strong gravitational lensing by rotating Simpson--Visser black holes,''
[arXiv:2104.00696 [gr-qc]].
%3 citations counted in INSPIRE as of 30 Aug 2021

%\cite{Tsukamoto:2020bjm}
\bibitem{Tsukamoto:2020bjm}
N.~Tsukamoto,
``Gravitational lensing in the Simpson-Visser black-bounce spacetime in a strong deflection limit,''
Phys. Rev. D \textbf{103}, no.2, 024033 (2021)
%doi:10.1103/PhysRevD.103.024033
[arXiv:2011.03932 [gr-qc]].
%8 citations counted in INSPIRE as of 30 Aug 2021

%\cite{Lobo:2020ffi}
\bibitem{Lobo:2020ffi}
F.~S.~N.~Lobo, M.~E.~Rodrigues, M.~V.~d.~S.~Silva, A.~Simpson and M.~Visser,
``Novel black-bounce spacetimes: wormholes, regularity, energy conditions, and causal structure,''
Phys. Rev. D \textbf{103}, no.8, 084052 (2021)
%doi:10.1103/PhysRevD.103.084052
[arXiv:2009.12057 [gr-qc]].
%15 citations counted in INSPIRE as of 30 Aug 2021

%\cite{Churilova:2019cyt}
\bibitem{Churilova:2019cyt}
M.~S.~Churilova and Z.~Stuchlik,
``Ringing of the regular black-hole/wormhole transition,''
Class. Quant. Grav. \textbf{37}, no.7, 075014 (2020)
%doi:10.1088/1361-6382/ab7717
[arXiv:1911.11823 [gr-qc]].
%16 citations counted in INSPIRE as of 30 Aug 2021



%\cite{Gundlach:1993tn}
\bibitem{Gundlach:1993tn}
C.~Gundlach, R.~H.~Price and J.~Pullin,
``Late time behavior of stellar collapse and explosions: 2. Nonlinear evolution,''
Phys. Rev. D \textbf{49}, 890-899 (1994)
%doi:10.1103/PhysRevD.49.890
[arXiv:gr-qc/9307010 [gr-qc]].
%153 citations counted in INSPIRE as of 09 Nov 2020

%\cite{Gundlach:1993tp}
\bibitem{Gundlach:1993tp}
C.~Gundlach, R.~H.~Price and J.~Pullin,
``Late time behavior of stellar collapse and explosions: 1. Linearized perturbations,''
Phys. Rev. D \textbf{49} (1994), 883-889
%doi:10.1103/PhysRevD.49.883
[arXiv:gr-qc/9307009 [gr-qc]].
%329 citations counted in INSPIRE as of 26 May 2020

%\cite{Leaver:1986gd}
\bibitem{Leaver:1986gd}
E.~W.~Leaver,
``Spectral decomposition of the perturbation response of the Schwarzschild geometry,''
Phys. Rev. D \textbf{34}, 384-408 (1986).
%doi:10.1103/PhysRevD.34.384
%274 citations counted in INSPIRE as of 09 Nov 2020

%\cite{Correa:2008nq}
\bibitem{Correa:2008nq}
D.~H.~Correa, J.~Oliva and R.~Troncoso,
``Stability of asymptotically AdS wormholes in vacuum against scalar field perturbations,''
JHEP \textbf{08}, 081 (2008)
%doi:10.1088/1126-6708/2008/08/081
[arXiv:0805.1513 [hep-th]].
%23 citations counted in INSPIRE as of 24 Nov 2021
	
%\cite{Evnin:2015gma}
\bibitem{Evnin:2015gma}
O.~Evnin and C.~Krishnan,
``A Hidden Symmetry of AdS Resonances,''
Phys. Rev. D \textbf{91} (2015) no.12, 126010
%doi:10.1103/PhysRevD.91.126010
[arXiv:1502.03749 [hep-th]].
%40 citations counted in INSPIRE as of 10 Nov 2021

%\cite{Evnin:2017vpc}
\bibitem{Evnin:2017vpc}
O.~Evnin, H.~Demirchian and A.~Nersessian,
``Mapping superintegrable quantum mechanics to resonant spacetimes,''
Phys. Rev. D \textbf{97}, no.2, 025014 (2018)
%doi:10.1103/PhysRevD.97.025014
[arXiv:1711.03297 [hep-th]].
%13 citations counted in INSPIRE as of 20 Nov 2021


%\cite{Fierro:2018rna}
\bibitem{Fierro:2018rna}
O.~Fierro, D.~Narbona, J.~Oliva, C.~Quijada and G.~Rubilar,
``Scalars on asymptotically locally AdS wormholes with $\mathcal{R}^{2}$ terms,''
[arXiv:1812.02089 [hep-th]].
%6 citations counted in INSPIRE as of 10 Nov 2021

%\cite{Anabalon:2019lzc}
\bibitem{Anabalon:2019lzc}
A.~Anabalon, J.~Oliva and C.~Quijada,
``Fully resonant scalars on asymptotically AdS wormholes,''
Phys. Rev. D \textbf{99} (2019) no.10, 104022
%doi:10.1103/PhysRevD.99.104022
[arXiv:1903.08239 [hep-th]].
%5 citations counted in INSPIRE as of 10 Nov 2021

%\cite{Shaikh:2021yux}
\bibitem{Shaikh:2021yux}
R.~Shaikh, K.~Pal, K.~Pal and T.~Sarkar,
``Constraining alternatives to the Kerr black hole,''
Mon. Not. Roy. Astron. Soc. \textbf{506}, no.1, 1229-1236 (2021)
%doi:10.1093/mnras/stab1779
[arXiv:2102.04299 [gr-qc]].
%13 citations counted in INSPIRE as of 20 Nov 2021



%\cite{Chatzifotis:2020oqr}
\bibitem{Chatzifotis:2020oqr}
N.~Chatzifotis, G.~Koutsoumbas and E.~Papantonopoulos,
``Formation of bound states of scalar fields in AdS-asymptotic wormholes,''
Phys. Rev. D \textbf{104}, no.2, 024039 (2021)
%doi:10.1103/PhysRevD.104.024039
[arXiv:2011.08770 [gr-qc]].
%4 citations counted in INSPIRE as of 16 Nov 2021

%\cite{Destounis:2019hca}
\bibitem{Destounis:2019hca}
K.~Destounis,
``Superradiant instability of charged scalar fields in higher-dimensional Reissner-Nordstr\"om-de Sitter black holes,''
Phys. Rev. D \textbf{100}, no.4, 044054 (2019)
%doi:10.1103/PhysRevD.100.044054
[arXiv:1908.06117 [gr-qc]].
%7 citations counted in INSPIRE as of 17 Nov 2020

%\cite{Babichev:2013cya}
\bibitem{Babichev:2013cya}
E.~Babichev and C.~Charmousis,
``Dressing a black hole with a time-dependent Galileon,''
JHEP \textbf{08} (2014), 106
%doi:10.1007/JHEP08(2014)106
[arXiv:1312.3204 [gr-qc]].
%242 citations counted in INSPIRE as of 24 Apr 2021

\end{thebibliography}
\end{document}